\documentclass[aps,prd,floatfix,12pt, notitlepage,a4paper,floatfix,showpacs,nofootinbib,superscriptaddress,tikz]{revtex4-2}
\pdfoutput=1
\RequirePackage[colorlinks=true,urlcolor=red,anchorcolor=red,citecolor=red,filecolor=red,linkcolor=red,menucolor=red,pagecolor=red,linktocpage=true,pdfproducer=medialab,pdfa=true]{hyperref}
\usepackage[usenames,dvipsnames]{color}  
\usepackage{graphicx}
\usepackage{adjustbox}
\usepackage{subcaption}
\usepackage{comment}
\usepackage[table]{xcolor} 
\usepackage{array} 
\usepackage{xcolor}
\usepackage{bm}
\usepackage{float}
\usepackage{bbold}
\usepackage{latexsym}
\usepackage{amsthm}
\usepackage{upgreek}
\usepackage{placeins}
\usepackage{wrapfig}
\usepackage{amssymb}
\usepackage{cancel}
\usepackage{multirow}
\usepackage{rotating}
\usepackage{geometry}
\usepackage{slashed}
\usepackage{subcaption}
\usepackage{hhline}
\usepackage{amsmath}
\usepackage{fancybox}
\usepackage[justification=raggedright,singlelinecheck=false]{caption}
\usepackage[utf8]{inputenc}
\usepackage[normalem]{ulem} 
\usepackage[normalem]{ulem}
\usepackage{color,soul}
\usepackage{tikz}
\usetikzlibrary{decorations.pathmorphing, arrows.meta}
\usetikzlibrary{decorations.markings}
\tikzset{
  fermion/.style={thick, postaction={decorate}, decoration={markings,mark=at position 0.5 with {\arrow{>}}}},
  scalar/.style={dashed, thick},
}

\setul{0.5ex}{0.3ex}
\setulcolor{blue}
\usepackage{braket}
\definecolor{linkcolor}{rgb}{0,0,0.5}
\allowdisplaybreaks
\allowdisplaybreaks[1]
\allowdisplaybreaks[2]

\definecolor{greenLinks}{rgb}{0, 0.6, 0}
\definecolor{blueLinks}{rgb}{0, 0, 0.6}
\definecolor{redLinks}{rgb}{0.6, 0, 0}
\definecolor{tempText}{rgb}{0.55, 0.10,0.67}
\definecolor{eprintLinks}{rgb}{0.4, 0.4, 0.4}
\definecolor{journalLinks}{rgb}{0.6, 0, 0}
\newcommand {\ignore}[1]{}
\usepackage{soul}
\definecolor{mightnightblue}{RGB}{25,25,112}
\definecolor{brown}{rgb}{0.59, 0.29, 0.0}
\definecolor{darkred}{rgb}{0.6,0,0}

\def\lsim{\mathrel{\rlap{\lower4pt\hbox{\hskip1pt$\sim$}}
    \raise1pt\hbox{$<$}}}
\def\gsim{\mathrel{\rlap{\lower4pt\hbox{\hskip1pt$\sim$}}
    \raise1pt\hbox{$>$}}}

\newcommand{\AddrBhopal}{Department of Physics, Indian Institute of Science Education and Research - Bhopal, Bhopal Bypass Road, Bhauri, Bhopal 462066, INDIA}

\begin{document}
\title{Twilight of the WIMP: Comprehensive Phenomenology of Electroweak Triplet Dark Matter}
\author{Gaurav}\email{gaurav24@iiserb.ac.in}
\affiliation{\AddrBhopal}
\author{Niharika Shrivastava}\email{niharikas21@iiserb.ac.in}
\affiliation{\AddrBhopal}
\author{Rahul Srivastava}\email{rahul@iiserb.ac.in}
\affiliation{\AddrBhopal}
\author{Sushant Yadav}\email{sushant20@iiserb.ac.in}
\affiliation{\AddrBhopal}
\begin{abstract}
\vspace{1cm}

We present a comprehensive study of dark matter phenomenology in standard model extensions featuring an electroweak triplet scalar or fermion with hypercharge 
$Y = 0$ or $Y = 2$. These minimal triplet extensions provide well-motivated dark matter candidates stabilised by the \(\mathbb{Z}_2\) discrete symmetry. We perform a detailed analysis of the parameter space consistent with current cosmological and experimental constraints, including the relic abundance, direct detection limits, and indirect detection bounds. We find that the scalar triplet with $Y=0$ is ruled out by a combination of relic density, direct detection and indirect detection constraints. On the other hand, the scalar and fermionic triplets with $Y=2$ are both excluded by current direct detection experiments due to their large spin-independent scattering cross-sections. The viable parameter space of the remaining $Y=0$ fermion triplet dark matter lies within the projected sensitivity of near-future experiments, particularly those targeting indirect detection signatures. Collider prospects for these triplet extensions are also discussed in the Appendix.
\end{abstract}

\maketitle 

\tableofcontents
\section{Introduction}

The Standard Model (SM), despite its many successes, lacks a viable dark matter (DM) candidate, pointing to new physics responsible for the observed relic abundance \cite{Planck:2018vyg}. Revealing the nature of DM is among the most profound unresolved questions in particle physics and cosmology. Despite overwhelming astrophysical and cosmological evidence, the particle nature and fundamental properties of DM remain unknown. Among the well-motivated candidates are weakly interacting massive particles (WIMPs), which can naturally reproduce the measured DM relic abundance via the thermal freeze-out mechanism \cite{Steigman:2012nb, Frumkin:2022}. Such a WIMP DM can be modelled by minimal extensions of the SM. One such minimal extension of the SM introduces a scalar singlet \cite{McDonald:1993ex,Bandyopadhyay:2010cc,Cline:2013gha} stabilised by a 
$\mathbb{Z}_2$ symmetry, providing a viable DM candidate. Another well-motivated realisation involves an additional electroweak scalar doublet whose neutral component serves as the DM candidate. Inert doublet DM has been widely explored in the context of the Scotogenic model \cite{Ma:2006km,Avila:2021mwg,Batra:2022pej} and its various extensions \cite{CentellesChulia:2019gic,Guo:2020qin,CentellesChulia:2022vpz,Kumar:2023moh,Kumar:2024zfb,CentellesChulia:2024iom,Singh:2025jtn,Kumar:2025cte,Kumar:2025zvv}. An equally compelling realisation of DM emerges in models with an inert scalar triplet \cite{FileviezPerez:2008bj,Araki:2011hm} through an appropriate extension of the SM gauge sector. Triplet DM is particularly appealing due to its predictive nature: the electroweak gauge and Higgs interactions largely determine its annihilation cross-sections, and its mass is constrained by relic density considerations. It also interacts readily with nuclei through gauge and Higgs portals. In addition, it leads to testable predictions for indirect detection and exhibits distinctive collider signatures, making it a highly testable and economical DM framework. \\
In this work, we focus on minimal extensions of the SM involving electroweak triplet fields, considering both scalar and fermion representations. We analyse two hypercharge assignments for each case: \(Y=0\) and \(Y=2\). These choices correspond to the minimal hypercharge assignments that allow the triplet to contain a neutral component suitable as a DM candidate. While the neutral components of these triplets serve as potential DM candidates, the charged components participate in electroweak interactions and can contribute to collider signatures. We perform a comprehensive phenomenological study of these models, including the calculation of the relic density, the evaluation of spin-independent and spin-dependent direct detection cross-sections and the computation of indirect detection signals arising from DM annihilation into SM particles, which can be probed through gamma-ray and cosmic-ray observations. \\
In addition, we examine the collider implications of these triplet fields. The charged and neutral components can be pair-produced via electroweak Drell–Yan processes and subsequently decay into lighter scalars or SM gauge bosons. Due to the presence of a stabilising $\mathbb{Z}_2$ symmetry, collider events are characterised by significant missing transverse energy alongside visible decay products, providing distinct signatures for current and future LHC searches.
This study thus provides a unified analysis of minimal triplet scalar and fermion DM models, exploring the interplay between cosmological constraints, direct and indirect detection, and collider phenomenology for both \(Y=0\) and \(Y=2\) scenarios. \\
This paper is organised as follows. In Sec.~\ref{sec:model}, we present four representative models of scalar and fermion triplets for both $Y=0$ and $Y=2$ scenarios. In Sec.~\ref{sec:constraints}, we discuss the various theoretical and experimental constraints applied to our models. In Sec.~\ref{sec:DM}, we discuss and compare the DM phenomenology for scalar and fermion triplet models for both 
$Y=0$ and $Y=2$ scenarios. The main conclusions of the paper are summarised in Sec.~\ref{sec:conclusions}. Finally, we have also included a brief discussion of the collider prospects of these triplet particles in Appendix~\ref{sec:Collider}.

\section{Model Framework}
\label{sec:model}

We begin by exploring two extensions of the SM, involving both scalar and fermion triplet fields that transform as triplets under the $SU(2)_L$ gauge group. Such multiplet extensions provide a minimal and well-motivated framework for addressing several open questions in particle physics, including the nature of DM and potential collider signatures beyond the SM. Depending on their hypercharge assignment, these triplet fields can lead to distinct phenomenological implications for relic density, direct and indirect detection, and collider observables. We consider $Y=0$ and $Y=2$ scalar or fermion triplets since these are the only minimal hypercharge assignments that contain a neutral component suitable for DM. \\
In particular, we consider the following four representative models:
\begin{itemize}
    \item \textbf{Scalar Triplet with Hypercharge $Y = 0$ ($\Delta_0$)}
    
    \item \textbf{Scalar Triplet with Hypercharge $Y = 2$ ($\Delta_2$)}
    
    \item \textbf{Fermion Triplet with Hypercharge $Y = 0$ ($\Sigma_0$)}
    
    \item \textbf{Fermion Triplet with Hypercharge $Y = 2$ ($\Sigma_2$)}
\end{itemize}
In the following subsections, we first discuss the scalar triplet extensions of the SM for both hypercharge assignments, followed by the fermionic triplet cases. The corresponding Lagrangian, particle spectrum, and relevant interaction terms are presented in detail, setting the stage for our subsequent analysis of DM phenomenology.

\subsection{Scalar Triplet with Hypercharge Y=0 ($\Delta_0$)}

We begin with the scalar triplet extension of the SM with hypercharge 
$Y=0$. We introduce a real scalar triplet \cite{Ross:1975fq,Forshaw:2001xq} $\Delta_0$, transforming as (1,3,0) under $SU(3)_C \otimes SU(2)_L \otimes U(1)_Y$ symmetry. The subscript ‘0’ in $\Delta_0$ denotes the hypercharge of the triplet. The neutral component, $\Delta_0^{0}$ serves as a viable DM candidate. To ensure its stability, we impose a \(\mathbb{Z}_2\) symmetry under which all SM fields are even, while the scalar triplet is odd. The corresponding field content and symmetry assignments are listed in Table~\ref{tab:delta0}. \\
The most general renormalisable Lagrangian for the scalar triplet with $Y=0$ can be written as
\begin{equation}
\mathcal{L}_{\Delta_0}  = (D_\mu H)^\dagger(D^\mu H) + \operatorname{Tr}(D_\mu \Delta_0)^\dagger(D^\mu \Delta_0) - V(H, \Delta_0).
\end{equation}
\begin{table}[H]
\centering
\begin{adjustbox}{max width=\textwidth}
\renewcommand{\arraystretch}{1.3} 
\begin{tabular}{|c||c|c|c||c|c||c|c|}
\hline
\multirow{2}{*}{\hspace{0.1cm}Gauge Group\hspace{0.1cm}} 
  & \multicolumn{3}{c||}{Baryon Fields} 
  & \multicolumn{2}{c||}{Lepton Fields} 
  & \multicolumn{2}{c|}{Scalar Fields} \\
\cline{2-8}
  & \hspace{0.1cm}$Q_L^i = (u_L^i, d_L^i)^T$\hspace{0.1cm} 
  & \hspace{0.1cm}$u_R^i$\hspace{0.1cm} 
  & \hspace{0.1cm}$d_R^i$\hspace{0.1cm} 
  & \hspace{0.1cm}$L_L^i = (\nu_L^i, e_L^i)^T$\hspace{0.1cm} 
  & \hspace{0.1cm}$e_R^i$\hspace{0.1cm} 
  & \hspace{0.1cm}$H$\hspace{0.1cm} 
  & \hspace{0.1cm}\textcolor{blue}{$\Delta_0$}\hspace{0.1cm} \\
\hline
\hline
$SU(3)_C$   & 3   & 3   & 3   & 1   & 1   & 1   & \textcolor{blue}{1}        \\
$SU(2)_L$   & 2   & 1   & 1   & 2   & 1   & 2& \textcolor{blue}{3}\\
$U(1)_Y$    & $1/6$ & $2/3$ & $-1/3$  &  $-1/2$ & $-1$ & $1/2$ & \textcolor{blue}{0}\\
$\mathbb{Z}_2$ & $+$ & $+$ & $+$ & $+$ & $+$ & $+$ & \textcolor{blue}{$-$} \\
\hline
\end{tabular}
\end{adjustbox}
\caption{Particle content and their corresponding charges under various symmetry groups, with the scalar triplet $\Delta_0$. Here the neutral component $\Delta_0^{0}$ of the triplet is the DM. }
\label{tab:delta0}
\end{table} 

The corresponding scalar potential takes the form
\begin{equation}
\begin{aligned}
V(H, \Delta_0) =\ & -\mu_H^2 H^\dagger H + \frac{1}{2}\mu_{\Delta_0}^2 \operatorname{Tr} (\Delta_0)^2 + \frac{1}{2}\lambda_H (H^\dagger H)^2 + \frac{1}{2} \lambda_\Delta \operatorname{Tr} (\Delta_0)^4 +  \\
& + \frac{1}{2}\lambda_{H\Delta} (H^\dagger H)\operatorname{Tr} (\Delta_0)^2.
\end{aligned}
\end{equation}
The gauge interactions of the triplet scalar are encoded in the covariant derivative, given by
\begin{equation} \label{eq:Covdev}
D_\mu \Delta_0= \partial_\mu \Delta_0 - i g [W_\mu, \Delta_0],
\end{equation}
where $g$ denotes the $SU(2)_L$ gauge coupling. The gauge field $W_\mu = \sum_{a=1}^{3} W_\mu^a T^a$, with $W_\mu^a$ and $T^a$ denote the $SU(2)_L$ gauge fields and generators, respectively. \\
Expanding the scalar fields in terms of their $SU(2)_L$ components, we can write 
\begin{equation}
H = \begin{pmatrix} 
G^+ \\ 
\frac{1}{\sqrt{2}} (v + h + i G^0) 
\end{pmatrix}, \quad
\Delta_0 = \begin{pmatrix}
\Delta^1\\
\Delta^2\\
\Delta^3
\end{pmatrix}.
\end{equation}
Here, $h$ is the SM Higgs boson and $G^{+}/G^{0}$ are the charged/neutral Goldstones. The parameter $v = 246.22$ GeV is the vacuum expectation value (VEV) of the Higgs field. \\
The real triplet field can be conveniently expressed in bi-doublet form. Using the pauli matrices $\sigma_i$, it can be written as $\Delta_0 = \frac{1}{\sqrt{2}}\sigma_i \Delta^i$,
where $\sigma_i$ are the Pauli matrices
$\sigma_1 = \begin{pmatrix}
0 & 1 \\
1 & 0
\end{pmatrix}, \quad
\sigma_2 = \begin{pmatrix}
0 & -i \\
i & 0
\end{pmatrix}, \quad
\sigma_3 = \begin{pmatrix}
1 & 0 \\
0 & -1
\end{pmatrix}.$ \\

Explicitly, this yields
\begin{equation}
\label{eq:adjrep}
\Delta_0 = \frac{1}{\sqrt{2}}
\begin{pmatrix}
\Delta^3 & \Delta^1-i\Delta^2 \\
\Delta^1+i\Delta^2 & -\Delta^3
\end{pmatrix}.
\end{equation}
In the charge eigenstate basis, the triplet components are identified as
\begin{equation}  
\Delta_0^0 \equiv{\Delta^3}, \quad \Delta_0^\pm = \frac{1}{\sqrt{2}}(\Delta^1 \mp i\Delta^2),
\end{equation}
such that the triplet field can be written in matrix form as
\begin{equation}
\Delta_0 =
\begin{pmatrix}
\frac{\Delta_0^0}{\sqrt{2}} & \Delta_0^+ \\
\Delta_0^- & -\frac{\Delta_0^0}{\sqrt{2}}
\end{pmatrix},
\end{equation}
where $\Delta_0^{\pm}$ and $\Delta_0^{0}$ represent the charged and neutral components of the scalar triplet field, respectively. \\
The neutral component of the scalar triplet field $\Delta_0^0$ is assumed not to acquire a VEV, i.e., $\langle \Delta_0^0 \rangle = 0$, in order to preserve the imposed \(\mathbb{Z}_2\) symmetry. A non-zero triplet VEV would spontaneously break the \(\mathbb{Z}_2\) symmetry and spoil the stability of the DM candidate \cite{Araki:2011hm}. After electroweak symmetry breaking, the Higgs boson and the components of the scalar triplet acquire masses, given by
\begin{equation}
m_h^2 = \lambda_H v^2,\quad
m_{\Delta_0^0}^2 = m_{\Delta_0^\pm}^2 = \mu_\Delta^2 + \frac{\lambda_{H\Delta} v^2}{2} =m_0^2.
\label{eq:massform}
\end{equation}
At tree level, the components of the scalar triplet are degenerate in mass, as no term in the scalar potential can create mass splitting. Since the neutral component of the scalar triplet field $\Delta_0^{0}$ does not acquire a VEV ($\langle \Delta_0^0 \rangle = 0$), electroweak symmetry breaking does not induce any additional mass splitting, and the components remain degenerate at this level. The mass splitting arises from electroweak radiative corrections, as loop diagrams involving electroweak gauge bosons contribute differently to the charged and neutral components of the triplet~\cite{Chiang:2020rcv,Cirelli:2005uq}. Mass splitting between the charged and the neutral components, depending on $m_0$ is given by
\begin{equation}
\Delta m = m_{\Delta_0^\pm} - m_{\Delta_0^0} = \frac{\alpha_w m_0}{4\pi}
\left[ f \left( \frac{m_W}{m_0} \right) - c_W^2 f \left( \frac{m_Z}{m_0} \right) \right],
\end{equation}
Here, $\alpha_{w} = \dfrac{g^2}{4\pi}$, $g$ denotes the gauge coupling constant associated with the $SU(2)_L$ weak interaction. $m_W$ and $m_Z$ are the masses of the $W$ and $Z$ bosons, respectively. $c_W = \cos\theta_W$ is the cosine of the weak mixing angle $\theta_W$.
The relevant loop functions can be expressed as~\cite{Chiang:2020rcv}
\begin{equation}
f(r) = -\frac{r}{4}[2 r^3 \log r - k r + (r^2 - 4)^{\frac{3}{2}} \ln A(r)],
\end{equation}
where $k$ denotes a UV regulator and,
\begin{equation}
A(r) = \frac{1}{2}(r^2-2-r \sqrt{r^2-2}).
\end{equation}
In the limit where $m_0$ is much greater than $m_W$, the above expression simplifies to $\Delta m \sim 160$ MeV~\cite{Araki:2011hm}.
This mass splitting means that the decay process $\Delta_0^{\pm} \to \Delta_0^{0} \pi^{\pm}$ is kinematically accessible, and the associated decay rate is given by~\cite{Chiang:2020rcv}
\begin{equation}
\Gamma\left(\Delta^{\pm} \to \Delta^0 \pi^{\pm}\right) = 
\frac{2 G_F^2}{\pi} \, f_\pi^2 V_{ud}^2 \, (\Delta m)^3 
\sqrt{1 - \frac{m_\pi^2}{(\Delta m)^2}}.
\end{equation}
The other parameters appearing in the expression are the Fermi constant \(G_F\), the pion decay constant \(f_{\pi} \approx 131\,\mathrm{MeV}\), 
the CKM matrix element \(V_{ud}\), and the pion mass \(m_{\pi}\). 
This particular decay channel is responsible for approximately 98\% of the total branching ratio, with the remaining decay modes being \(\Delta_0^{\pm} \to \Delta_0^{0} \mu^{\pm} \nu_{\mu}\) and \(\Delta_0^{\pm} \to \Delta_0^{0} e^{\pm} \nu_{e}\). As a consequence, the charged scalar particle exhibits a relatively long lifetime, approximately \(\tau_{\Delta_0^{\pm}} \sim 0.17~\mathrm{ns}\).


\subsection{Scalar Triplet with Hypercharge Y=2 ($\Delta_2$)}

 We now consider a modified version of the previous model in which the scalar triplet carries hypercharge $Y=2$. Accordingly, we denote the triplet field as $\Delta_2$, where the subscript indicates its hypercharge assignment. Due to this choice of hypercharge, the three components of the triplet possess electric charges 0, +1, and +2, respectively. As in the $Y=0$ case, we impose a discrete $\mathbb{Z}_2$ symmetry under which the triplet field is odd while all SM fields are even, thereby ensuring the stability of the DM candidate. The particle content, along with their corresponding charges, is summarised in Table~\ref{tab:scalarY2}.
\begin{table}[h!]
\centering
\begin{adjustbox}{max width=\textwidth}
\renewcommand{\arraystretch}{1.3} 
\begin{tabular}{|c||c|c|c||c|c||c|c|}
\hline
\multirow{2}{*}{\hspace{0.1cm}Gauge Group\hspace{0.1cm}} 
  & \multicolumn{3}{c||}{Baryon Fields} 
  & \multicolumn{2}{c||}{Lepton Fields} 
  & \multicolumn{2}{c|}{Scalar Fields} \\
\cline{2-8}
  & \hspace{0.1cm}$Q_L^i = (u_L^i, d_L^i)^T$\hspace{0.1cm} 
  & \hspace{0.1cm}$u_R^i$\hspace{0.1cm} 
  & \hspace{0.1cm}$d_R^i$\hspace{0.1cm} 
  & \hspace{0.1cm}$L_L^i = (\nu_L^i, e_L^i)^T$\hspace{0.1cm} 
  & \hspace{0.1cm}$e_R^i$\hspace{0.1cm} 
  & \hspace{0.1cm}$H$\hspace{0.1cm} 
  & \hspace{0.1cm}\textcolor{blue}{$\Delta_2$}\hspace{0.1cm} \\
\hline
\hline
$SU(3)_C$   & 3   & 3   & 3   & 1   & 1   & 1   & \textcolor{blue}{1}        \\
$SU(2)_L$   & 2   & 1   & 1   & 2   & 1   & 2& \textcolor{blue}{3}\\
$U(1)_Y$    & $1/6$ & $2/3$ & $-1/3$  &  $-1/2$ & $-1$ & $1/2$ & \textcolor{blue}{2}\\
$\mathbb{Z}_2$ & $+$ & $+$ & $+$ & $+$ & $+$ & $+$ & \textcolor{blue}{$-$} \\
\hline
\end{tabular}
\end{adjustbox}
\caption{Particle content and their corresponding charges under various symmetry groups, with the scalar triplet $\Delta_2$.}
\label{tab:scalarY2}
\end{table}
\\
The Lagrangian for the scalar sector is given by 
\begin{equation}
\mathcal{L}_{\Delta_2} = (D_\mu H)^\dagger(D^\mu H) + \operatorname{Tr}(D_\mu \Delta_2)^\dagger(D^\mu \Delta_2) - V(H, \Delta_2).
\end{equation}
The corresponding scalar potential can be written as
\begin{equation}
\begin{aligned}
V(H, \Delta_2) =\ & -\mu^2 H^\dagger H + \mu_{\Delta_2}^2 \operatorname{Tr} (\Delta_2^{\dagger}\Delta_2) + \frac{1}{2}\lambda_H (H^\dagger H)^2 + \frac{1}{2} \lambda_{\Delta} \operatorname{Tr} (\Delta_2^{\dagger}\Delta_2\Delta_2^{\dagger}\Delta_2)+  \\
& + \frac{1}{2}\lambda^{\prime}_{\Delta} (\operatorname{Tr}[\Delta_2^{\dagger}\Delta_2])^2 + \frac{1}{2} \lambda_{H\Delta} H^\dagger H \operatorname{Tr}[{\Delta_2^{\dagger}\Delta_2}] + \frac{1}{2} \lambda^{\prime}_{H\Delta} H^\dagger{\Delta_2\Delta_2^{\dagger}} H .
\end{aligned}
\end{equation}
The covariant derivative for scalar triplet $\Delta_2$ is given by
\begin{equation} \label{eq:Covdev2}
D_\mu \Delta_2= \partial_\mu \Delta_2 - i g [W_\mu, \Delta_2]-2 i g^{\prime}B_{\mu}\Delta_2,
\end{equation}
here, $g$ and $g^{\prime}$ denote the gauge couplings corresponding to the $SU(2)_L$ and $U(1)_Y$ gauge groups, respectively. The gauge field $W_\mu = \sum_{a=1}^{3} W_\mu^a T^a$ denotes the $SU(2)_L$ gauge fields and $B_{\mu}$ is the $U(1)_Y$ gauge field. $H$ denotes the SM Higgs doublet, and the scalar triplet $\Delta_2$ is expressed as
\begin{equation}
\Delta_2 = \begin{pmatrix}
\Delta_2^1\\
\Delta_2^2\\
\Delta_2^3
\end{pmatrix}.
\end{equation}
Expanding in the bi-doublet representation, as in Eq.~\eqref{eq:adjrep}, we obtain

\begin{equation}
\label{eq:adjrep2}
\Delta_2 = \frac{1}{\sqrt{2}}
\begin{pmatrix}
\Delta^3_2 & \Delta^1_2-i\Delta^2_2 \\
\Delta_2^1+i\Delta_2^2 & -\Delta_2^3
\end{pmatrix},
\end{equation}
where the component fields are identified according to their electric charges as
\begin{equation}  
\Delta^+_2\equiv\Delta_2^3, \quad \Delta_2^{++} = \frac{\Delta_2^1-i\Delta_2^2}{\sqrt{2}},  \quad \Delta_2^0= \frac{\Delta_2^1+i\Delta_2^2}{\sqrt{2}}.
\end{equation}
In terms of the charge eigenstates, the triplet can be written in the bi-doublet representation as

\begin{equation}
\Delta_2 =
\begin{pmatrix}
\frac{\Delta_2^+}{\sqrt{2}} & \Delta_2^{+ +}\\
\Delta_2^0 & -\frac{\Delta^+_2}{\sqrt{2}}
\end{pmatrix}.
\end{equation}
After electroweak symmetry breaking, the mass spectrum of the Higgs boson and the components of the scalar triplet are obtained as
\begin{eqnarray}\label{eq:masses}
m_h^2 & = & \lambda_H v^2,  \nonumber \\
m_{\Delta_{2}^0}^2
& = & \mu_{\Delta_2}^2 + \tfrac{1}{4}(\lambda_{H\Delta} + \lambda^{\prime}_{H\Delta})v^2, \nonumber \\
m_{\Delta_2^\pm}^2 
& = & \mu_{\Delta_2}^2 + \tfrac{1}{4}\!\left(\lambda_{H\Delta} + \tfrac{\lambda^{\prime}_{H\Delta}}{2}\right)v^2, \nonumber \\
m_{\Delta_2^{\pm\pm}}^2 
& = & \mu_{\Delta_2}^2 + \tfrac{1}{4}\lambda_{H\Delta} v^2.
\end{eqnarray}
 In contrast to the $Y=0$ case, the charged and neutral components in the $Y=2$ scenario are non-degenerate in mass. This is because the additional quartic interactions in the scalar potential induce mass splittings among the triplet components after electroweak symmetry breaking. The neutral component will be our DM candidate, and in order to ensure that it is the lightest $\mathbb{Z}_2$-odd particle, we take $\lambda'_{H\Delta} < 0$ throughout this work.

\subsection{Fermion Triplet with Hypercharge Y=0 ($\Sigma_0$)}

 We next consider the fermion triplet extension of the SM. In this case, the SM is extended by the addition of a Majorana fermion triplet $\Sigma_0$, transforming under $SU(3)_C \otimes SU(2)_L \otimes U(1)_Y$ as (1,3,0). Additionally, a discrete \(\mathbb{Z}_2\) symmetry is imposed, under which the fermion triplet is odd while all SM fields are even. The particle content of the model, along with their corresponding quantum numbers, is summarised in Table~\ref{tab:ferm0}.
\begin{table}[h!]
\centering
\begin{adjustbox}{max width=\textwidth}
\renewcommand{\arraystretch}{1.3}
\begin{tabular}{|c||c|c|c||c|c||c||c|}
\hline
\multirow{2}{*}{\hspace{0.1cm}Gauge Group\hspace{0.1cm}} 
  & \multicolumn{3}{c||}{Baryon Fields} 
  & \multicolumn{2}{c||}{Lepton Fields} 
  & \multicolumn{1}{c||}{Scalar Fields}
  & \multicolumn{1}{c|}{Fermion Fields} \\
\cline{2-8}
  & \hspace{0.1cm}$Q_L^i = (u_L^i, d_L^i)^T$\hspace{0.1cm} 
  & \hspace{0.1cm}$u_R^i$\hspace{0.1cm} 
  & \hspace{0.1cm}$d_R^i$\hspace{0.1cm} 
  & \hspace{0.1cm}$L_L^i = (\nu_L^i, e_L^i)^T$\hspace{0.1cm} 
  & \hspace{0.1cm}$e_R^i$\hspace{0.1cm} 
  & \hspace{0.1cm}$H$\hspace{0.1cm}
  & \hspace{0.1cm}\textcolor{blue}{$\Sigma_0$}\hspace{0.1cm} \\
\hline
\hline
$SU(3)_C$   & 3   & 3   & 3   & 1   & 1   & 1   & \textcolor{blue}{1}     \\
$SU(2)_L$   & 2   & 1   & 1   & 2   & 1   & 2   & \textcolor{blue}{3}     \\
$U(1)_Y$    & $1/6$ & $2/3$ & $-1/3$ & $-1/2$ & $-1$ & $1/2$ & \textcolor{blue}{$0$} \\
$\mathbb{Z}_2$ & $+$ & $+$ & $+$ & $+$ & $+$ & $+$ & \textcolor{blue}{$-$} \\
\hline
\end{tabular}
\end{adjustbox}
\caption{Particle content and their corresponding charges under various symmetry groups, including the fermion triplet $\Sigma_0$. The neutral component of $\Sigma_0$ serves as the DM candidate.}
\label{tab:ferm0}
\end{table}
\\
The Lagrangian of this framework is given as
\begin{equation}
\mathcal{L}_{\Sigma_0} = \frac{1}{2}\mathrm{Tr} \left[ \overline{\Sigma_0}\, i\gamma^\mu D_\mu \Sigma_0 \right]  -\frac{1}{2} M_{\Sigma}~Tr(\overline{\Sigma_0} \Sigma_0),
\end{equation}
where $M_{\Sigma}$ represents the Majorana mass parameter of the fermion triplet $\Sigma_0$. The covariant derivative has the same form as in Eq.~\ref{eq:Covdev}. \\
The fermion triplet $\Sigma_0$ can be expressed in the bi-doublet representation as
\begin{equation}
\Sigma_0 = 
\begin{pmatrix}
\frac{\Sigma_0^0}{\sqrt{2}} & \Sigma_0^+ \\
\Sigma_0^- & -\frac{\Sigma_0^0}{\sqrt{2}}
\end{pmatrix}
\end{equation}
In addition to the neutral component $\Sigma_0^0$, the triplet contains charged fermions $\Sigma_0^\pm$ with electric charges of $\pm$1. Unlike the scalar  cases, here the annihilation and co-annihilation processes involving the neutral component $\Sigma_0^0$ and its charged counterparts $\Sigma_0^\pm$ are primarily mediated only by the SM gauge bosons. The imposed $\mathbb{Z}_2$ symmetry forbids the renormalisable Yukawa interaction  $\bar{L}_L \tilde{H}\Sigma_0 + h.c.$, which would otherwise induce mixing between the lepton doublets and the fermion triplet. As a consequence, the triplet fermion cannot decay into SM particles, ensuring the stability of the DM candidate. 


\subsection{Fermion Triplet with Hypercharge Y=2 ($\Sigma_2$)}
 In analogy with the scalar triplet scenario, we consider a fermion triplet with hypercharge $Y=2$. Unlike the $Y=0$ case, this representation is complex, requiring the introduction of vector-like fermions $\Sigma_{2L}$ and $\Sigma_{2R}$. The particle content and details on additional \(\mathbb{Z}_2\) charge assignment are shown in Table~\ref{table:fermionY=2}.
\begin{table}[h!]
\centering
\begin{adjustbox}{max width=\textwidth}
\renewcommand{\arraystretch}{1.3}
\begin{tabular}{|c||c|c|c||c|c||c||c|c|}
\hline
\multirow{2}{*}{\hspace{0.1cm}Gauge Group\hspace{0.1cm}} 
  & \multicolumn{3}{c||}{Baryon Fields} 
  & \multicolumn{2}{c||}{Lepton Fields} 
  & \multicolumn{1}{c||}{Scalar Fields}
  & \multicolumn{2}{c|}{Fermion Fields} \\
\cline{2-9}
  & \hspace{0.1cm}$Q_L^i = (u_L^i, d_L^i)^T$\hspace{0.1cm} 
  & \hspace{0.1cm}$u_R^i$\hspace{0.1cm} 
  & \hspace{0.1cm}$d_R^i$\hspace{0.1cm} 
  & \hspace{0.1cm}$L_L^i = (\nu_L^i, e_L^i)^T$\hspace{0.1cm} 
  & \hspace{0.1cm}$e_R^i$\hspace{0.1cm} 
  & \hspace{0.1cm}$H$\hspace{0.1cm}
  & \hspace{0.1cm}\textcolor{blue}{$\Sigma_{2L}$}\hspace{0.1cm}
  & \hspace{0.1cm}\textcolor{blue}{$\Sigma_{2R}$}\hspace{0.1cm} \\
\hline
\hline
$SU(3)_C$   & 3   & 3   & 3   & 1   & 1   & 1   & \textcolor{blue}{1} & \textcolor{blue}{1} \\
$SU(2)_L$   & 2   & 1   & 1   & 2   & 1   & 2   & \textcolor{blue}{3} & \textcolor{blue}{3} \\
$U(1)_Y$    & $1/6$ & $2/3$ & $-1/3$ & $-1/2$ & $-1$ & $1/2$ & \textcolor{blue}{$2$} & \textcolor{blue}{$2$} \\
$\mathbb{Z}_2$ & $+$ & $+$ & $+$ & $+$ & $+$ & $+$ & \textcolor{blue}{$-$} & \textcolor{blue}{$-$} \\
\hline
\end{tabular}
\end{adjustbox}
\caption{Particle content and their corresponding charges under various symmetry groups, including the fermion triplet fields $\Sigma_{2L}$ and $\Sigma_{2R}$.}
\label{table:fermionY=2}
\end{table}
\\
The Lagrangian of this framework is given as

\begin{equation}
\mathcal{L}_{\Sigma_2} = \mathrm{Tr} \left[ \overline{\Sigma_2}\, i\gamma^\mu D_\mu \Sigma_2 \right] -  M_{\Sigma} \ Tr(\overline{{\Sigma_{2}}} \Sigma_{2}),
\end{equation}
where $M_{\Sigma}$ denotes the Dirac mass parameter of the fermion triplet. The covariant derivative has the same form as in Eq.~\ref{eq:Covdev2}. \\
The fermion triplet $\Sigma_2$ is a Dirac field defined as
\begin{equation}
    \Sigma_2 = \Sigma_{2L} + \Sigma_{2R},
\end{equation}
with $\Sigma_{2L}$ and $\Sigma_{2R}$ transforming identically as (1,3,2) under $SU(3)_C \otimes SU(2)_L \otimes U(1)_Y$. In the bi-doublet representation, it can be written as
\begin{equation}
\Sigma_{2} = 
\begin{pmatrix}
\frac{\Sigma_2^+}{\sqrt{2}} & \Sigma_2^{++} \\
\Sigma_2^0 & -\frac{\Sigma_2^+}{\sqrt{2}}
\end{pmatrix},
\end{equation}
where the component fields carry electric charges of 0, +1, and +2, respectively. As a consequence of its non-zero hypercharge, the fermion triplet couples to both charged and neutral gauge bosons, leading to interactions with $W^{\pm}$, $Z$ and the photon. At tree level, all components of the fermion triplet are degenerate in mass, with $m_{\Sigma_2^0} = m_{\Sigma_2^{\pm}} = m_{\Sigma_2^{\pm\pm}}=M_{\Sigma}$. Radiative corrections lift this degeneracy, generating small mass splittings among the components. In particular, the singly and doubly charged states receive positive mass shifts of approximately 160 MeV and 675 MeV, respectively, relative to the neutral component $\Sigma_2^0$ \cite{McKay:2017xlc}. Consequently, the neutral component remains the lightest state and serves as the DM candidate in our analysis.

\section{Constraints on the Model}
\label{sec:constraints}
 Prior to discussing the DM phenomenology, we summarise the theoretical and experimental constraints that govern the viable parameter space of the model. All numerical results are obtained after consistently implementing these constraints.

\subsection{Perturbativity and Boundedness from below}{\label{subsec:BFB}}
We require all couplings to be perturbative and impose the following condition.
\begin{equation}
\lambda_H,~\lambda_{\Delta},~\lambda_{H\Delta} \leq 4\pi.
\end{equation}
The stability of the scalar potential at tree level requires the following conditions among the potential parameters.
For the Y=0 scalar triplet case, the constraints are given by \cite{Araki:2011hm}
\begin{equation}
\lambda_H>0, \quad \lambda_{\Delta}>0, \quad \lambda_{H\Delta} > -2\sqrt{\lambda_H \lambda_\Delta}.
\end{equation}
 In contrast, for the Y=2 scalar triplet scenario, the stability conditions take the form \cite{Arhrib:2011uy}

\begin{align}
& \lambda_H > 0, \lambda_{\Delta} + \lambda^{\prime}_{\Delta} > 0, \lambda^{\prime}_{\Delta} + \frac{\lambda_{\Delta}}{2} > 0,\nonumber \\
& \lambda_{H\Delta} + \sqrt{2\,\lambda_H (\lambda_{\Delta} + \lambda^{\prime}_{\Delta})} > 0,\lambda_{H\Delta} + \sqrt{2\,\lambda_H \left(\lambda^{\prime}_{\Delta} + \frac{\lambda_{\Delta}}{2}\right)} > 0, \nonumber \\
& \lambda_{H\Delta} + \lambda^{\prime}_{H\Delta} + \sqrt{2\,\lambda_H (\lambda_{\Delta} + \lambda^{\prime}_{\Delta})} > 0,\lambda_{H\Delta} + \lambda^{\prime}_{H\Delta} + \sqrt{2\,\lambda_H \left(\lambda^{\prime}_{\Delta} + \frac{\lambda_{\Delta}}{2}\right)} > 0. \nonumber \\
\end{align}
For the fermionic scenarios, since no extra scalar fields are present, the scalar potential coincides with the SM Higgs potential, and thus the standard SM perturbativity and stability conditions apply.
\subsection{Electroweak precision data}
 Electroweak precision observables provide stringent constraints on the parameter space of the model. In particular, the oblique parameters 
$S$, $T$, and $U$ encapsulate the radiative effects of heavy new states on the gauge boson self-energies. The analytical expressions for the oblique parameters $S$, $T$, and $U$ for both $Y=0$ and $Y=2$ triplet scalars are provided in Ref.~\cite{Cheng:2022hbo}. In our numerical scan, the values of 
$S$, $T$, and $U$ are computed using SPheno-4.0.5 \cite{Porod:2011nf}, and the resulting predictions are required to satisfy the current experimental limits
\cite{ParticleDataGroup:2024cfk}
\begin{eqnarray}
S &=& -0.04 \pm 0.10, \\
T &=& 0.01 \pm 0.12, \\
U &=& -0.01 \pm 0.09.
\end{eqnarray}

\subsection{Constraints from Higgs Invisible Decay}

 We impose the constraint on the branching ratio of the Higgs invisible decay into dark sector particles based on the most recent LHC data \cite{ATLAS:2023tkt}. The BSM channel that contributes to invisible Higgs decays for the two scalar DM cases are $h \rightarrow \Delta^{0}_{0} \Delta^{0}_{0}$, $h \rightarrow \Delta^{0}_{2} \Delta^{0}_{2}$. So, we are imposing 
\begin{equation}
\text{BR}(h \rightarrow \Delta^{0}_{0} \Delta^{0}_{0})
, \text{BR}(h \rightarrow \Delta^{0}_{2} \Delta^{0}_{2})
< 0.107
\end{equation}
For fermion triplet DM with Y=0 or Y=2, a gauge and $\mathbb{Z}_2$ symmetry-invariant renormalisable Yukawa interaction with the SM Higgs doublet cannot be constructed. Therefore, the Higgs does not couple to the neutral triplet fermion at tree level, and the Higgs invisible decay into these dark sector particles is absent.

\subsection{LEP Constraints}
\label{subsec:LEP}
LEP provides the most precise measurement of the invisible decay width of the $Z$-boson \cite{LEP:2000pgt,Carena:2003aj,ALEPH:2005ab,OPAL:2003wxm,OPAL:2003nhx}. In models with electroweak scalar or fermion triplets, additional contributions to the invisible $Z$-boson decay may arise depending on the hypercharge assignment of the triplet. For a real triplet with $Y=0$, the neutral component carries $T_3=0$ and therefore does not couple directly to the $Z$-boson at tree level. As a result, the invisible decay constraint from $Z$-pole measurements is naturally avoided in this case. \\
On the other hand, for triplets with nonzero hypercharge, such as a scalar or fermion triplet with $Y=2$, the neutral component possesses gauge interactions with the $Z$-boson. Consequently, processes such as
$Z \rightarrow \Delta_2^0 \Delta_2^{0*}$
(or the corresponding fermionic channel) can contribute to the invisible decay width whenever kinematically allowed. To satisfy the stringent LEP bounds on the invisible $Z$-boson width, these decay channels must therefore be forbidden, typically requiring the triplet states to be heavier than $m_Z/2$. \\
Additionally, LEP searches for charged scalar particles impose important bounds on the scalar triplet parameter space. These constraints arise primarily from direct pair production at LEP-II through electroweak interactions. For the $Y=0$ scalar triplet, the spectrum consists of a neutral component $\Delta_0^0$ and charged states $\Delta_0^\pm$. The charged scalars can be pair-produced via $e^+e^- \to \gamma^*/Z \to \Delta_0^+\Delta_0^-$. The subsequent signatures involve missing energy due to the stable neutral component $\Delta_0^0$, leading to constraints similar to those from chargino searches. As a result, the charged scalar mass is constrained to be $m_{\Delta_0^\pm} \gtrsim 100~\text{GeV}\,$. For the $Y=2$ scalar triplet, the spectrum includes $\Delta_2^{++}$, $\Delta_2^+$, and $\Delta_2^0$. In addition to singly charged scalar production, the presence of doubly charged scalars leads to distinctive signatures. In particular, pair production processes such as $e^+e^- \to \gamma^*/Z \to \Delta_2^{++}\Delta_2^{--}$ provide strong constraints. LEP searches for doubly charged scalars impose lower bounds of approximately
$m_{\Delta_2^{\pm\pm}} \gtrsim 100~\text{GeV}\,$, with comparable bounds on the singly charged states depending on the decay channels~\cite{Heister:2002mn,Abbiendi:2003pr,Achard:2003mv}. In both cases, constraints from the $Z$-boson width are generally subdominant for scalar triplets due to the suppressed production near threshold, and the dominant limits arise from direct searches at LEP-II. \\
We now discuss the LEP bounds on fermionic electroweak triplets. For the $Y=0$ fermion triplet, the spectrum consists of $\Sigma_0^0$ and $\Sigma_0^\pm$, with the neutral component being the lightest. Due to its vanishing weak isospin and hypercharge, the neutral state does not couple to the $Z$-boson at tree level. Consequently, the decay $Z \to \Sigma_0^0 \Sigma_0^0$ is absent. However, the charged components couple to the electroweak gauge bosons and can be pair-produced via $e^+e^- \to \gamma^*/Z \to \Sigma_0^+ \Sigma_0^-$. The absence of such contributions to the precisely measured $Z$-boson width implies the kinematic constraint $2\,m_{\Sigma_0^\pm} > m_Z\,$.
In addition, $W$-boson decays such as $W^{\pm} \to \Sigma_0^\pm \Sigma_0^0$ require
$m_{\Sigma_0^\pm} + m_{\Sigma_0^0} > m_W\,$. Stronger bounds arise from direct searches at LEP-II. Since the charged fermions couple to gauge bosons analogously to wino-like charginos, one can apply the LEP limits on chargino production, leading to $m_{\Sigma_0^\pm} > 103.5~\text{GeV}\,,$
as reported in Refs.~\cite{ALEPH:2003acj,LEPChargino:2003}.
For the $Y=2$ fermion triplet, the spectrum contains $\Sigma_2^{++}$, $\Sigma_2^+$, and $\Sigma_2^0$. In this case, the neutral component has a non-vanishing coupling to the $Z$-boson, and all components participate in electroweak interactions. As a result, LEP-I measurements of the $Z$-boson width constrain the decays, $Z \to \Sigma_2^0 \Sigma_2^0$, $Z \to \Sigma_2^+ \Sigma_2^-$, $Z \to \Sigma_2^{++} \Sigma_2^{--}\,$, which leads to the bounds $2\,m_{\Sigma_2^0} > m_Z$,
$2\,m_{\Sigma_2^+} > m_Z,$
$2\,m_{\Sigma_2^{++}} > m_Z\,$. Furthermore, kinematic constraints from $W$-boson decays imply $m_{\Sigma_2^0} + m_{\Sigma_2^+} > m_W$,
$m_{\Sigma_2^+} + m_{\Sigma_2^{++}} > m_W\,$. Direct searches at LEP-II for charged fermions provide the most stringent limits. Pair production processes such as $e^+e^- \to \gamma^*/Z \to \Sigma_2^+ \Sigma_2^-$ and $\Sigma_2^{++} \Sigma_2^{--}$ lead to lower bounds of approximately
$m_{\Sigma_2^+} \gtrsim 103.5~\text{GeV}\,$, $m_{\Sigma_2^{++}} \gtrsim 100~\text{GeV}\,,$
with the precise values depending on the decay modes~\cite{ALEPH:2003acj,vonderPahlen:2016cbw}.

\subsection{LHC constraints}
\label{lhc}
At the LHC, triplet dark sector particles are constrained mainly through electroweak production of charged components. The limits therefore arise from searches for charged tracks, disappearing tracks, and multi-lepton signatures. For the scalar triplet with $Y=0$, radiative corrections induce a small mass splitting between the charged and neutral components of $\Delta m \sim 160$ MeV. In this case, the charged scalar predominantly decays via $\Delta^\pm_0 \rightarrow \Delta^0_0 \pi^\pm$, producing a soft pion and a disappearing charged track signature at the LHC. Searches for long-lived charged particles using disappearing tracks constrain such electroweak triplet states, requiring the charged scalar mass to be roughly $m_{\Delta^\pm_0} \gtrsim 300\text{--}350~\mathrm{GeV}$~\cite{ATLAS:2022rme}. For the scalar triplet with $Y=2$, the spectrum contains a doubly charged scalar $\Delta^{\pm\pm}_2$ in addition to $\Delta^\pm_2$ and $\Delta^0_2$. At the LHC, these states can be produced through electroweak processes such as $pp \rightarrow \Delta^{++}_2\Delta^{--}_2$ and $pp \rightarrow \Delta^{\pm\pm}_2\Delta^\mp_2$. The doubly charged scalar leads to striking collider signatures through decays like $\Delta^{\pm\pm}_2\rightarrow \ell^\pm \ell^\pm$ or $\Delta^{\pm\pm}_2\rightarrow W^\pm W^\pm$. Searches for same-sign dilepton final states at the LHC place strong bounds on the mass of the doubly charged scalar, typically requiring $m_{\Delta^{\pm\pm}_2} \gtrsim 800$--$900~\mathrm{GeV}$ depending on the branching ratios~\cite{ATLAS:2017xqs}. \\
For the fermion triplet with $Y=0$, radiative corrections induce a small mass splitting between the charged and neutral components of the triplet, given by $\Delta m = m_{\Sigma^\pm_0} - m_{\Sigma^0_0} \simeq 160~\text{MeV}$~\cite{Cirelli:2005uq}. Consequently, the charged component predominantly decays via $\Sigma^\pm_0 \to \Sigma^0_0 \pi^\pm$ through an off-shell $W^\pm$ boson, leading to a decay length of $\mathcal{O}(\text{cm})$. Such a long-lived charged particle results in a disappearing track signature at the LHC. Searches for disappearing tracks by the ATLAS and CMS collaborations constrain wino-like electroweak fermion triplets and exclude masses up to approximately $400$--$430~\text{GeV}$ at $95\%$ CL~\cite{ATLAS:2019gqq,CMS:2018rea}. For the fermion triplet with $Y=2$, the multiplet contains the states $\Sigma^{++}_2$, $\Sigma^{+}_2$ and $\Sigma^{0}_2$, where the neutral component $\Sigma^{0}_2$ can serve as the DM candidate. These states are pair-produced at the LHC through electroweak gauge interactions mediated by $\gamma$, $Z$ and $W^\pm$. The heavier charged components decay through gauge interactions via the cascade, $\Sigma^{\pm\pm}_2\to\Sigma^{\pm}_2W^{\pm}$ followed by $\Sigma^{\pm}_2\to\Sigma^{0}_2W^{\pm}$, resulting in final states with multiple $W$-bosons and missing transverse energy. Searches for electroweak multiplets with charged states at the LHC constrain such scenarios through multilepton and gauge boson-rich signatures. Current analyses exclude fermion triplet masses below approximately $300-350~\text{GeV}$, depending on the mass spectrum and decay topology~\cite{delAguila:2008cj,Franceschini:2008pz}.


\section{Dark Matter Phenomenology}
\label{sec:DM}
We now focus on the analysis of the DM. In the early universe, all particles in the dark sector remain in thermal equilibrium with the SM plasma through a set of production and annihilation processes. As the universe expands and cools, the interaction rates eventually drop below the Hubble expansion rate, rendering the thermal bath incapable of efficiently producing heavy and unstable dark-sector states. These particles subsequently decay or annihilate into lighter particles. By contrast, the lightest particle in the dark sector is stabilised by the discrete $\mathbb{Z}_2$
symmetry and is therefore cosmologically stable. Once this stable particle decouples from the thermal bath, its co-moving number density ceases to evolve, leading to the freeze-out of its relic abundance.
The resulting DM relic density can then be  computed and compared with Planck observations \cite{Planck:2018vyg}, which require 0.1126 $\leq \Omega h^2 \leq$ 0.1246 at the 3$\sigma$ confidence level\cite{Batra:2023bqj}. The production-annihilation diagrams used for computation of relic density are shown in Appendix~\ref{sec:appendix}. \\
The DM in our model could also be detected by direct detection experiments such as XENONnT \cite{XENON:2025vwd}, LZ \cite{LZ:2022lsv,LZ:2024zvo} and PandaX-4T \cite{PandaX:2024qfu} through the Higgs and $Z$ portals, leading to additional constraints on the parameter space. The relevant diagrams are shown in Fig.~\ref{fig:ddTS0} in Appendix~\ref{sec:appendix}. The DM candidate can also be probed via indirect detection experiments such as Fermi-LAT~\cite{Fermi-LAT:2016uux}, H.E.S.S.~\cite{HESS:2022ygk}, VERITAS~\cite{VERITAS:2024usn}, ANTARES and ARCA~\cite{KM3NeT:2024xca}, through observations of gamma rays and neutrinos arising from the dominant annihilation channel $\text{DM DM} \rightarrow W^+ W^-$, thereby providing complementary constraints on the model parameter space. The relevant diagrams are shown in Fig.~\ref{fig:ddTS01}. Detailed discussions are presented in Secs.~\ref{subsubsec:IDST} and \ref{subsubsec:IDFT} for the scalar and
fermion triplet cases, respectively.
The models are implemented in Feynrules 2.3 \cite{Alloul:2013bka}. The DM relic density, DM-nucleon scattering cross-sections, and indirect detection rates are determined by micrOMEGAS 6.3.0 \cite{Belanger:2006is,Belanger:2014vza,Alguero:2023zol}. We begin with the discussion of the scalar triplets in \ref{subsec:scalartriplet} while the fermionic triplets are discussed later in \ref{subsec:fermiontriplet}.
\subsection{Scalar Triplet DM} \label{subsec:scalartriplet}

 We now discuss the DM phenomenology of the scalar triplet.  In the scalar triplet model with $Y=0(2)$, the neutral component of the triplet ($\Delta_0^0(\Delta_2^0)$) serves as the DM candidate. The scalar potential of our models contains several dimensionless quartic coupling parameters and one dimensionful mass parameter, which we vary over the following range in Table~\ref{tab:param_ranges}.
\begin{table}[h!]
\centering
\begin{subtable}{0.45\textwidth}
\centering
\captionsetup{justification=centering}
\begin{tabular}{|c|c|}
\hline
\textbf{Parameter} & \textbf{Range} \\
\hline\hline
$\lambda_H$& [0.2586] \\
\hline
$\lambda_\Delta$& $[10^{-8}, 4\pi]$ \\
\hline
$\lambda_{H\Delta}$& $ \pm[10^{-8}, 4\pi]$ \\
\hline
$\mu_{\Delta_0}^2$ & $[10^0, 3 \times 10^4]^2$~$\text{GeV}^2$ \\
\hline
\end{tabular}
\caption{Scalar triplet with $Y=0$}
\end{subtable}
\hspace{1cm}
\begin{subtable}{0.45\textwidth}
\centering
\captionsetup{justification=centering}
\begin{tabular}{|c|c|}
\hline
\textbf{Parameter} & \textbf{Range} \\
\hline\hline
$\lambda_H$& $[0.2586]$ \\
\hline
$\lambda_\Delta$& $\pm[10^{-8}, {4\pi}]$ \\
\hline
$\lambda^{\prime}_\Delta$& $\pm[10^{-8}, {4\pi}]$ \\
\hline
$\lambda_{H\Delta}$& $ \pm[10^{-8}, {4\pi}]$ \\
\hline
$\lambda^{\prime}_{H\Delta}$& $-[10^{-8}, {4\pi}]$ \\
\hline
$\mu_{\Delta_2}^2$ & $[10^0, 3 \times 10^4]^2$~$\text{GeV}^2$ \\
\hline
\end{tabular}
\caption{Scalar triplet with $Y=2$}
\end{subtable}
\caption{Parameter space ranges used in the scan for scalar triplets. See text for the detailed discussion regarding the choice of the range.}
\label{tab:param_ranges}
\end{table} 
\\
The Higgs quartic coupling $\lambda_H = 0.2586$ is fixed in both cases, determined at the tree level by Eqn.~\eqref{eq:massform} consistent with the measured value of Higgs mass $m_h \approx 125.20$ GeV~\cite{ParticleDataGroup:2024cfk}.
The triplet self-coupling $\lambda_\Delta \in [10^{-8}, 4\pi]$ (for $Y=0$) is restricted to positive values to satisfy the bounded-from-below conditions as mentioned in section \ref{subsec:BFB}. The lower bound $\lambda_\Delta \geq 10^{-8}$ is imposed, as further decreasing its value does not lead to any significant variation in the computed observables within numerical accuracy. The upper bound $\lambda_\Delta \leq 4\pi$ is set by perturbative unitarity~\cite{Krauss:2017xpj}. The Higgs portal coupling $\lambda_{H\Delta}$ is varied in the range $\pm[10^{-8}, 4\pi]$. It can take both positive and negative values, as it does not affect the stability of the scalar potential but instead contributes directly to the physical mass of the DM candidate, while remaining consistent with bounded-from-below and perturbativity constraints. \\
For the $Y = 2$ case, the scalar triplet sector contains two additional couplings $\lambda'_\Delta$ and $\lambda'_{H\Delta}$ compared to the $Y=0$ scenario. Unlike the $Y = 0$ case, where $\lambda_\Delta$ must be strictly positive to satisfy the bounded-from-below conditions, in the $Y=2$ scenario neither $\lambda_\Delta$ nor $\lambda'_\Delta$ is required to be individually positive. Instead, the stability of the scalar potential is ensured by the condition $\lambda_\Delta + \lambda'_\Delta > 0$, as discussed in Sec.~\ref{subsec:BFB}. This relaxes the individual constraints, allowing both $\lambda_\Delta$ and $\lambda'_\Delta$ to vary over the range $\pm[10^{-8}, 4\pi]$. Similarly, both Higgs portal couplings $\lambda_{H\Delta}$ and $\lambda'_{H\Delta}$ are allowed to take positive as well as negative values. They do not individually affect the stability of the scalar potential but instead contribute to the physical masses and mass splittings among the neutral and charged components of the complex triplet spectrum. \\
The mass-squared parameter is varied in the range $1 \leq \mu_{\Delta_0}^2, \mu_{\Delta_2}^2 \leq (3 \times 10^4)^2$ GeV$^2$. The upper bound is chosen as no viable relic density satisfying the observed limits is obtained beyond this range in our numerical scan. We now discuss the relic density of the scalar triplet DM.

\subsubsection{Relic Density} \label{subsubsec:relicST}

\begin{figure}[H]
    \centering
    \includegraphics[width=1\linewidth]{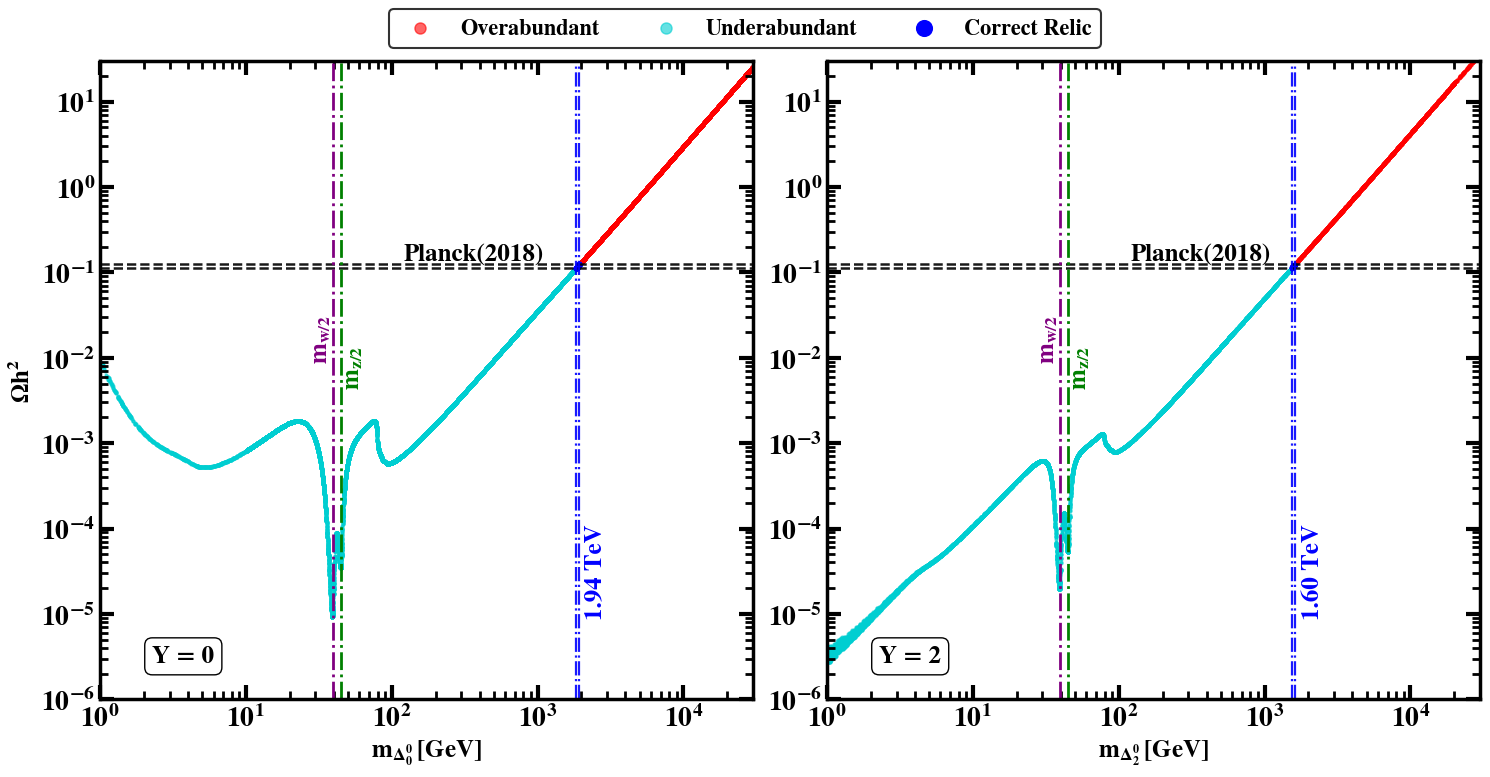}
    \caption{Relic density of the DM candidate as a function of the DM mass in the scalar triplet model with negligible Higgs portal coupling. The left (right) panel corresponds to the hypercharge $Y=0$ ($Y=2$) case.}
    \label{fig:SRDsmall}
\end{figure}

 We present the relic density $\Omega h^{2}$ of the scalar triplet DM candidate as a function of its mass. Results are shown separately for the two hypercharge assignments of the scalar
triplet: $Y = 0$ (left panels) and $Y = 2$ (right panels). The dark-turquoise, red,
and blue points represent under-abundant,
over-abundant, and correct relic density, respectively~\cite{Planck:2018vyg}. 
Fig.~\ref{fig:SRDsmall} displays the relic density in the limit of negligible Higgs portal coupling, such that the phenomenology is governed solely by gauge interactions of the triplet with the $W$ and $Z$ bosons. 
 This choice isolates the pure gauge contribution to the relic density and enables a direct assessment of how the $SU(2)_{L} \otimes U(1)_{Y}$ quantum numbers, together with the DM mass, govern the cosmological abundance through gauge interactions, independent of Higgs-mediated effects. \\
In the low-mass region, the relic density decreases rapidly with increasing mass due to enhanced annihilation via gauge bosons. Two pronounced dips are visible in both panels. The first dip appears near \(m_W/2 \simeq 40.2\,\text{GeV}\), where the coannihilation cross-section is enhanced due to resonant \(W\)-mediated processes, leading to a sharp decrease in \(\Omega h^{2}\), see Fig.~\ref{fig:CoannihiTS2_y0} and Fig.~\ref{fig:CoannihiTS2}. The second dip, located near $m_Z/2 \simeq 45.6\,\text{GeV}$, have  distinct 
dynamical origins for the $Y=0$ and $Y=2$ cases. For the $Y=0$ case, as the mass of the nearly degenerate charged components 
$\Delta_0^{\pm}$ approaches half the $Z$-boson mass, the coannihilation process 
$\Delta^{+}_0\Delta^{-}_0 \to Z^* \to \bar{q}q\,/\,l^+l^-$ (as illustrated in 
Fig.~\ref{fig:CoannihiTS2_y0}) becomes resonantly enhanced, leading to a sharp increase 
in the effective annihilation cross-section and a corresponding suppression 
of the relic abundance. For the $Y=2$ case, the dip is because of the resonant annihilation of DM through s-channel $Z$-boson mediated processes as shown in Fig.~\ref{fig:annihiTS2}. Notably, for both cases, the two dips occur at different depths, with the $W$-resonance dip consistently deeper than the $Z$-resonance dip. This asymmetry arises because the triplet couples to the $W$-boson through the full $SU(2)_{L}$ gauge coupling $g$, while the coupling to the $Z$-boson is suppressed by
the weak mixing angle. 
In addition, coannihilation channels mediated by the $Z$-boson act indirectly to enhance the effective cross-section.
Consequently, the $W$-mediated processes lead to a larger
annihilation rate, producing a deeper resonance dip. 

Beyond the $Z$ boson mass, the relic density increases with the DM mass, reaching the Planck's observed value at $m_{\Delta_0^0} \simeq 1.94~\text{TeV}$ for the $Y=0$ scalar triplet and at $m_{\Delta_{2}^0} \simeq 1.60~\text{TeV}$ for the $Y=2$ case. We obtain the correct relic density only within a very narrow mass region due to the absence of the Higgs portal interactions. We now include the Higgs portal coupling and examine the resulting changes in the relic density plot. 

\begin{figure}[h!]
    \centering
    \includegraphics[width=1\linewidth]{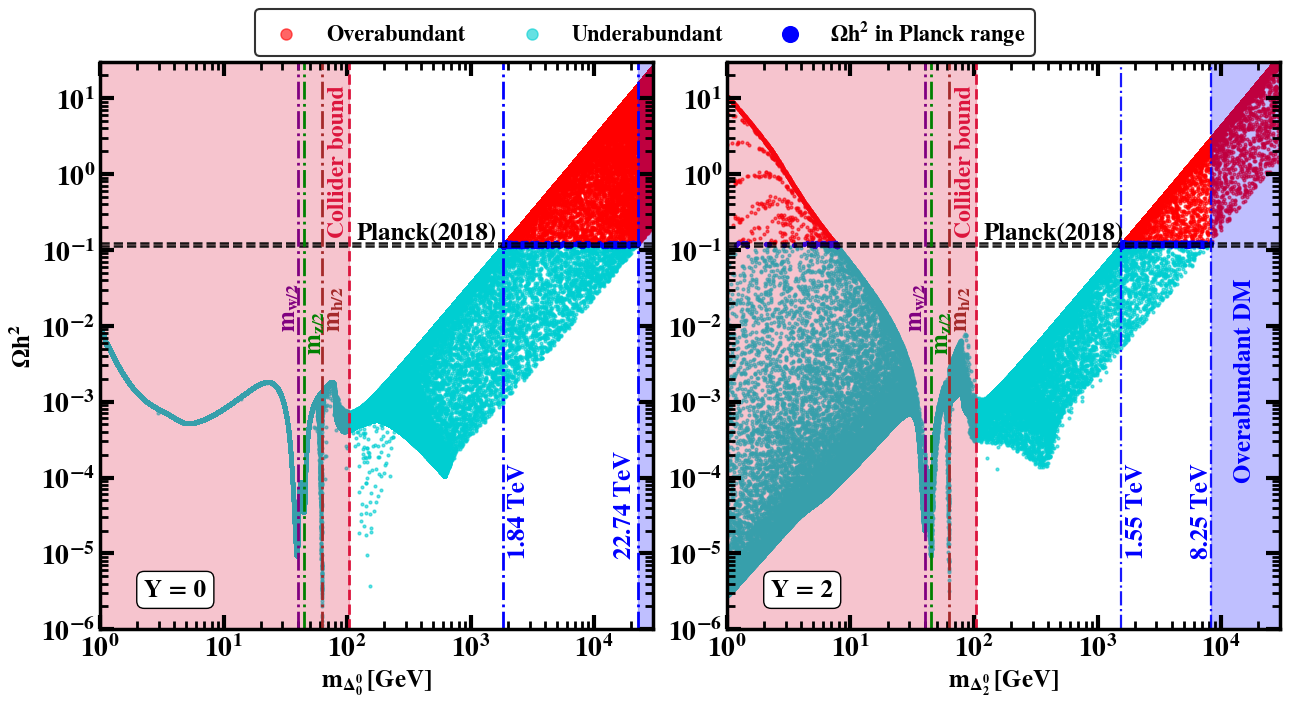}
    \caption{Relic density of the DM candidate as a function of the DM mass in the scalar triplet model with finite Higgs portal coupling.}
    \label{fig:SRD}
\end{figure}

\begin{figure}[H]
    \centering
    \includegraphics[width=1\linewidth]{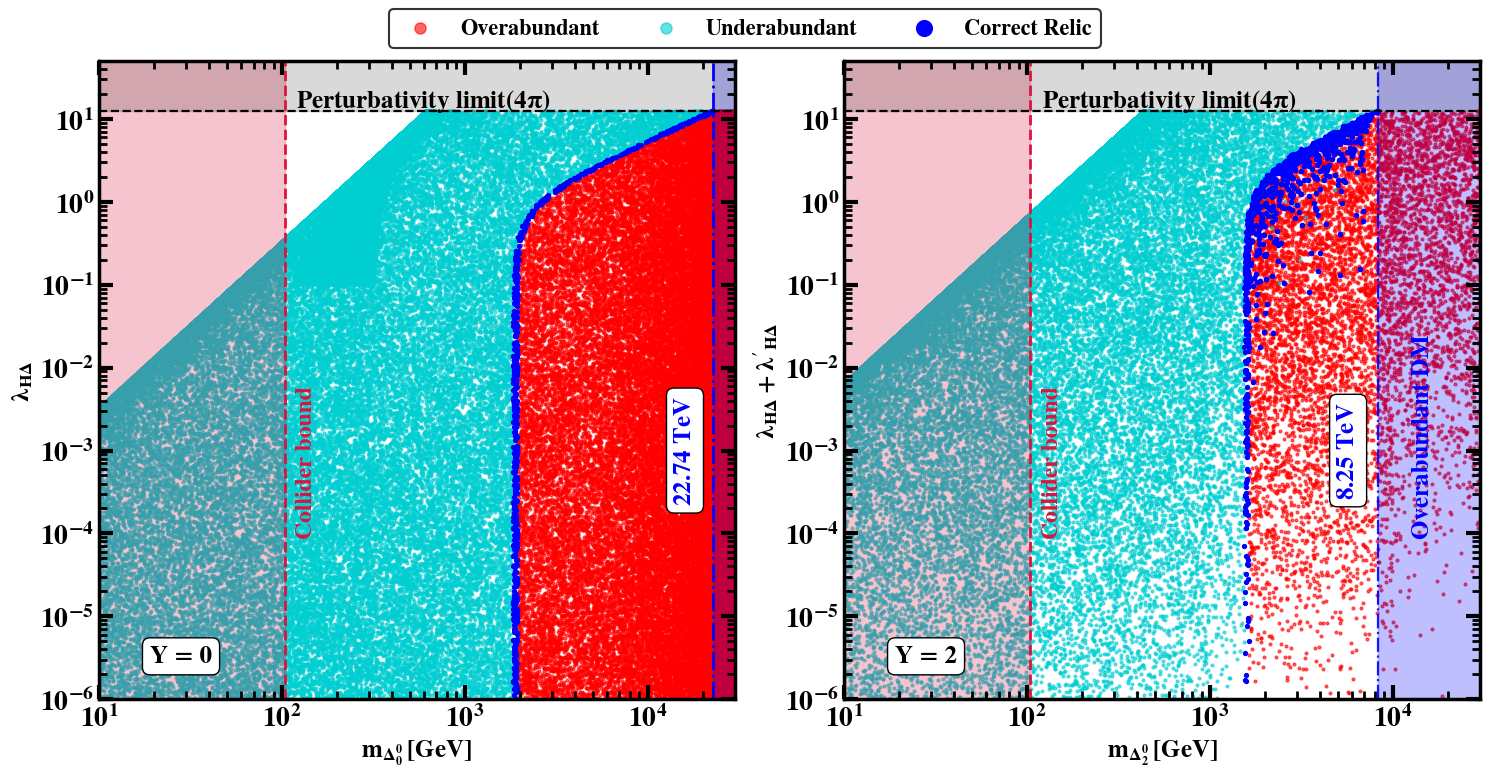}
    \caption{This figure shows the variation of cross-coupling between Higgs-DM with the mass of DM candidate. Here, the color scheme used is same as above. The vertical shaded lines correspond to the collider bound and only overabundent DM region, respectively. The horizontal shaded region corresponds to the tree level perturbativity limit on the couplings.}
    \label{fig:lam_mass}
\end{figure}

Fig.~\ref{fig:SRD} extends the analysis to the case of finite Higgs portal
couplings, with all relevant
parameters varied within the ranges specified in Table~\ref{tab:param_ranges}. The inclusion of a
finite $\lambda_{H\Delta}$ and $\lambda'_{H\Delta}$ introduces a third resonance: a new dip appears at
$m_{h}/2 \simeq 62.6$~GeV, arising from the $s$-channel
Higgs resonance in the annihilation process. 
The dark-turquoise points below the Planck band correspond to regions where DM annihilation is too efficient, leading to a relic density smaller than the observed value \cite{Planck:2018vyg}. These points are not excluded but indicate that the scalar triplet contributes only a fraction of the total DM. The red points correspond to the overabundant region, where annihilation is inefficient and the relic density exceeds the observed value. These points predict $\Omega h^{2} > 0.1246$ and are therefore inconsistent with the Planck measurement unless additional mechanisms reduce the DM abundance. The blue points correspond to regions that reproduce the observed relic abundance, satisfying the Planck constraint on $\Omega h^{2}$ \cite{Planck:2018vyg}. A
vertical dashed line at $m_{\Delta_{0(2)}^0} \simeq 103.5$~GeV marks the lower bound on
the charged scalar mass derived from LEP searches for pair-produced exotic
charged particles, as detailed in Section~\ref{subsec:LEP}.  All parameter
points to the left of this boundary are excluded by direct collider searches,
regardless of their relic density value. \\
At high DM masses, the Planck-compatible region forms a broad band rather than a thin strip, bounded by two approximately parallel edges. These boundaries arise from the variation of the Higgs portal coupling. The upper edge corresponds to larger values of $\lambda_{H\Delta}$/$\lambda^{\prime}_{H\Delta}$, where enhanced Higgs-mediated annihilation requires a higher DM mass to reproduce the observed relic density. In contrast, the lower edge corresponds to negligible portal coupling, approaching the pure gauge limit of Fig.~\ref{fig:SRDsmall}. The finite width of this band reflects the sensitivity of the relic density to the Higgs portal interaction, which becomes increasingly pronounced at high masses. Comparing the $Y = 0$ and $Y = 2$ cases reveals clear qualitative differences. The Planck-compatible region is broader in the $Y = 0$ case and narrower for $Y = 2$. 
For the $Y = 0$ triplet, the DM is a real scalar whose annihilation is governed by the Higgs and $SU(2)_L$ gauge interactions, leading to a relatively milder dependence of the annihilation cross-section on the DM mass in the TeV range. As a result, the observed relic density can be achieved over a wider mass range.
In contrast, for $Y = 2$, the presence of hypercharge enhances the overall annihilation rate and introduces additional channels. This leads to a stronger mass dependence of the annihilation cross-section, so the relic density matches the observed value within a narrower range of DM masses, resulting in a thinner allowed band. \\
A further distinction appears in the low-mass region of Fig.~\ref{fig:SRD}. For $Y = 2$, the relic density distribution exhibits a visible wide range which is much less pronounced in the $Y = 0$ case. This feature arises due to the presence of two independent Higgs portal couplings, $\lambda_{H\Delta}$ and $\lambda^{\prime}_{H\Delta}$, which contribute differently to the annihilation cross-section. Depending on their relative values and signs, the Higgs-mediated contribution can either enhance or partially suppress the total annihilation rate near the Higgs resonance, leading to the observed wide range.
In contrast, for $Y = 0$, only a single Higgs portal coupling is present, and the annihilation is dominated by gauge interactions. As a result, the Higgs contribution plays a subdominant role and a narrow band is observed. 
After imposing all experimental constraints, a sizable region of parameter space remains consistent with the observed relic density. For the $Y=0$ case, the allowed DM mass spans $m_{\Delta^0_0} \simeq 1.84-22.74~\text{TeV}$, while for $Y=2$ it lies in the range $m_{\Delta^0_{2}} \simeq 1.55-8.25~\text{TeV}$. \\
To further elaborate the role of the Higgs portal, in Fig.~\ref{fig:lam_mass} we show the variation of the Higgs-DM coupling as a function of the DM mass for both hypercharge assignments, using the same color convention as before. The shaded region at large coupling values indicates the perturbativity limit. The vertical dashed line represents the collider bound, excluding the low-mass region $m_{\Delta_{0(2)}^0} \lesssim 103.5~\text{GeV}$ independent of the coupling. 
In the $Y = 0$ case (left panel), the Planck-compatible blue points form a diagonal band in the $(\lambda_{H\Delta},\,m_{\Delta^0_0})$ plane, starting around $m_{\Delta^0_0} \sim 1.8$--$1.9~\text{TeV}$ for small Higgs portal couplings and extending up to $m_{\Delta^0_0} \simeq 22.74~\text{TeV}$ at larger couplings near the perturbativity limit. This positive correlation indicates that larger Higgs portal couplings are required at higher DM masses to reproduce the observed relic density. At large masses, the gauge interactions alone are not sufficient, and the Higgs-mediated contribution becomes important. The lower edge of the blue band, where the coupling approaches zero around $m_{\Delta^0_0} \sim 1.8~\text{TeV}$, corresponds to the pure gauge limit, where gauge interactions alone yield the correct relic abundance, consistent with Fig.~\ref{fig:SRDsmall}. As the Higgs portal coupling increases towards its perturbative limit, the correct relic density can be satisfied at higher DM masses, reaching up to $m_{\Delta^0_0} \simeq 22.74~\text{TeV}$ near the perturbativity bound. 
A qualitatively similar structure is observed for the $Y = 2$ case (right
panel), but with two notable differences.  First, the $Y = 2$ triplet scalar
possesses two independent Higgs portal couplings, $\lambda_{H\Delta}$ and
$\lambda^{\prime}_{H\Delta}$, so the quantity plotted on the vertical axis is
their sum $\lambda_{H\Delta}+ \lambda^{\prime}_{H\Delta}$ which also explains the wider distribution of correct relic density points in the high-mass region.  The perturbativity bound on this combined coupling is accordingly twice as large, $4\pi$ on each, which extends the theoretically accessible range of the vertical axis and permits Planck-compatible solutions up to $m_{\Delta_2^0} \simeq 8.25$~TeV, a considerably
lower upper limit than the $22.74$~TeV reached in the $Y = 0$ scenario.  This
reduction is a direct consequence of the enhanced gauge annihilation rate of the $Y = 2$ representation: because the gauge interactions already provide a larger contribution to the total cross-section, the Higgs portal coupling needs to compensate less at any given mass, and the perturbativity limit is encountered at a lower DM mass than in the $Y = 0$ case. It is also important to note that in Fig.~\ref{fig:lam_mass} both Higgs portal couplings were varied within the range 
\([10^{-8},\,4\pi]\). However, since \(\lambda'_{H\Delta} < 0\), the combination 
\(\lambda_{H\Delta} + \lambda'_{H\Delta}\) is bounded from above by the perturbativity limit of \(4\pi\).
 Second, the blue Planck-compatible strip in the $Y = 2$ panel is noticeably steeper and more compressed in the
mass direction, consistent with the narrower viable mass range observed in
Fig.~\ref{fig:SRD}. Furthermore, there are isolated scattered blue points in $Y=2$ plot, which reflect the fact that there are two different Higgs-DM couplings at play  and the relic can be satisfied by their different combinations of values. Taken together, Fig.~\ref{fig:lam_mass} demonstrates that the correct relic density can be achieved across several orders of magnitude in the Higgs portal coupling for both hypercharge assignments, with the viable mass range determined by the interplay between the gauge coupling strength, the Higgs portal contribution, the perturbativity and collider constraints. We now focus on the direct detection prospects of scalar triplet DM.
\subsubsection{Direct Detection} \label{subsubsec:DDST}

\begin{figure}[H]
    \centering
    \includegraphics[width=1\linewidth]{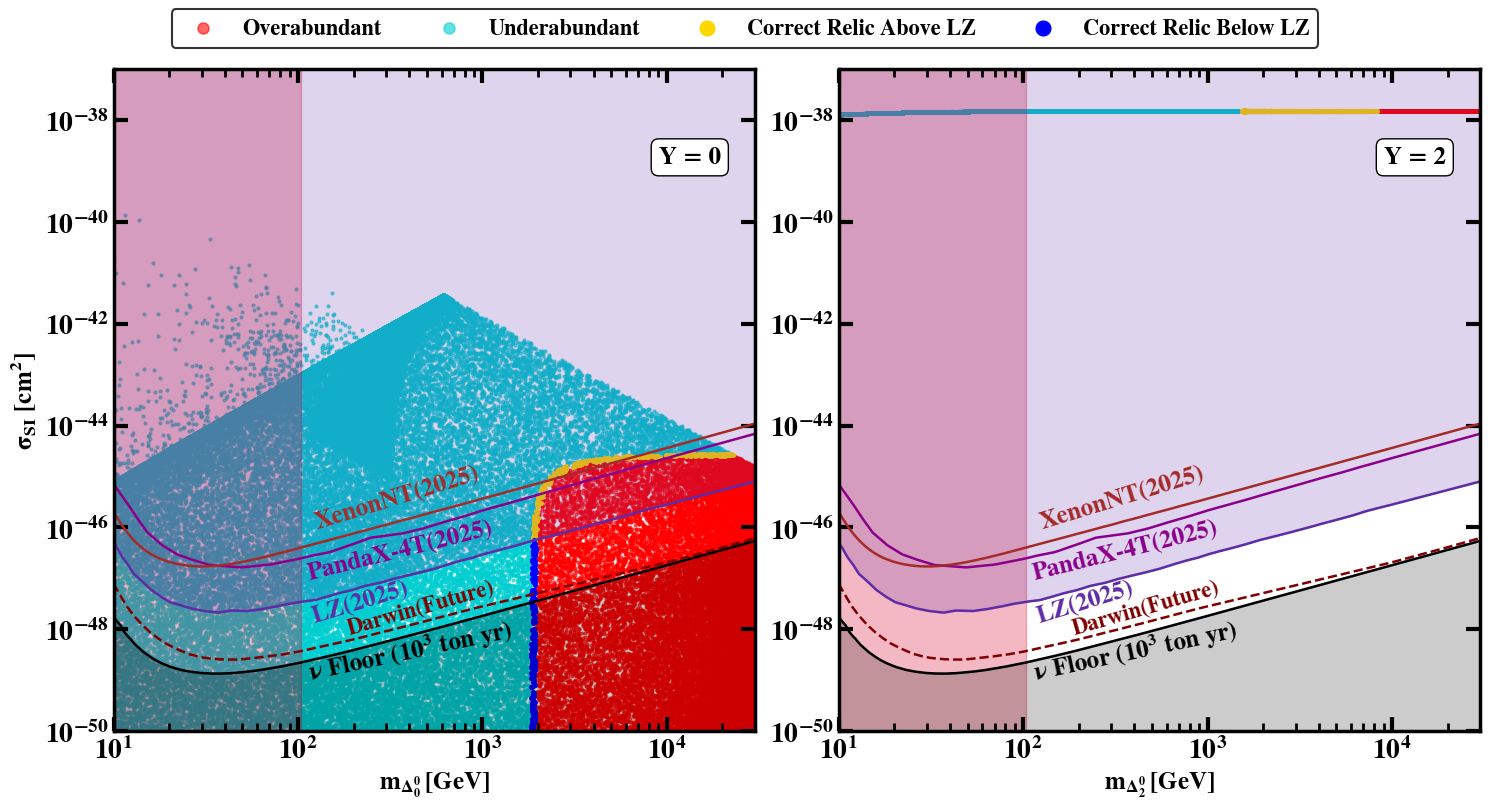}
    \caption{\small The spin-independent WIMP-nucleon cross-section for the scalar triplet DM candidate is shown as a function of DM mass. The color coding follows the same scheme as described earlier, gold points lying above the current LZ (2025) exclusion limit~\cite{LZ:2024zvo} are ruled out. The black shaded region indicates the neutrino floor~\cite{AristizabalSierra:2021kht, DeRomeri:2025nkx}, below which direct detection experiments become progressively insensitive due to irreducible neutrino backgrounds.}
    \label{fig:DD_Scalar}
\end{figure}

 After identifying the region of parameter space compatible with the observed relic density, we investigate the implications for direct detection experiments. Direct Detection experiments aim to observe the rare elastic scattering of galactic DM particles off nuclei in terrestrial detectors, measuring the small recoil energies ($\sim$keV) deposited in these interactions. Fig.~\ref{fig:DD_Scalar} displays the spin-independent WIMP-nucleon scattering cross-section $\sigma_{\text{SI}}$ as a function of the scalar triplet DM mass, following the same color convention as before. The relevant Feynman diagrams contributing to direct detection for the $Y=0$ ($Y=2$) scenarios are presented in Appendix~\ref{sec:appendix} in Fig.~\ref{fig:ddTS0}. \\
The exclusion limits from current and projected direct detection experiments are also shown: LZ~2025~\cite{LZ:2024zvo} (purple), PandaX-4T~\cite{PandaX:2024qfu} (dark magenta), XENONnT~2025~\cite{XENON:2025vwd} (maroon), and DARWIN (future)~\cite{DARWIN:2016hyl} (brown). The neutrino floor is indicated by the black curve.
Parameter points excluded by the current LZ bound are highlighted in gold, while points lying below the neutrino floor are shown in  a black shaded region. Notably, the phenomenologically viable region satisfying the Planck relic density (blue points) lies largely within the projected sensitivity of upcoming direct detection experiments for both hypercharge assignments. \\
After imposing the LZ constraint, the relic density compatible DM survives in a very narrow mass range of
$1.56$--$1.94~\mathrm{TeV}$ for $Y=0$. In the low-mass region, the points located above the boundary predominantly arise from parameter space characterised by the Higgs portal coupling varying over a broad range.
The outer boundary of the scatter plot corresponds to the extremal values of the Higgs–DM cross-coupling. Moving from the lower to the upper edge of the plot, $\lambda_{H\Delta}$ increases monotonically and eventually saturates at $4\pi$. Beyond the apex of this boundary, the behavior becomes asymmetric: towards lower masses, the coupling decreases continuously, reaching values as small as $\mathcal{O}(10^{-8})$, whereas towards higher masses it remains fixed at its maximal value up to the final over-abundant points. Along the over-abundant boundary, $\lambda_{H\Delta}$ decreases progressively as one moves from the highest DM masses towards smaller values.
Towards the extreme high-mass end of the parameter space, the number of Planck-compatible points decreases rapidly.
This behaviour originates from the fact that the Higgs–portal couplings approach their maximal values, significantly affecting both the annihilation cross-section governing the relic density and the elastic scattering amplitude relevant for direct detection.
As a consequence, the allowed parameter space becomes increasingly compressed, and the predicted direct detection signal strength diminishes toward the edge of the scan. \\
For the complex scalar triplet with $Y = 2$, the DM-nucleon spin-independent 
scattering cross-section receives a tree-level contribution from $Z$-boson 
exchange due to the non-zero hypercharge of the DM candidate. This 
yields $\sigma_{\rm SI} \sim 10^{-38}~\text{cm}^2$, which exceeds the current 
LZ exclusion limit by several orders of magnitude, thereby ruling out the 
entire parameter space of this model \cite{YaserAyazi:2015tfq,Araki:2011hm}. 
In summary, the $Y=0$ DM case still has allowed, albeit a small, parameter space left after imposing the relic density and direct detection constraints along with all the other constraints mentioned in Sec.~\ref{sec:constraints}. On the other hand, the entire parameter space of the $Y=2$ DM is ruled out by the direct detection constraints, implying that DM cannot be a $Y=2$ $SU(2)_L$ triplet scalar.
We now discuss the indirect detection prospects of scalar triplet DM.

\subsubsection{Indirect Detection} \label{subsubsec:IDST}

Indirect detection experiments probe SM particles produced via DM annihilation in astrophysical environments with high DM density, such as the Galactic Center, dwarf spheroidal galaxies, and the Sun. The resulting SM particles propagate to terrestrial detectors and may appear as an excess in the flux of cosmic messengers over the expected astrophysical background. The primary indirect detection probes include high-energy $\gamma$-rays, neutrinos, and charged cosmic rays such as positrons, antiprotons, and antinuclei. The dominant Feynman diagrams relevant for the indirect detection of the scalar triplet are shown in Fig.~\ref{fig:ddTS01}. \\
\begin{figure}[H]
    \centering
    \includegraphics[width=1\linewidth]{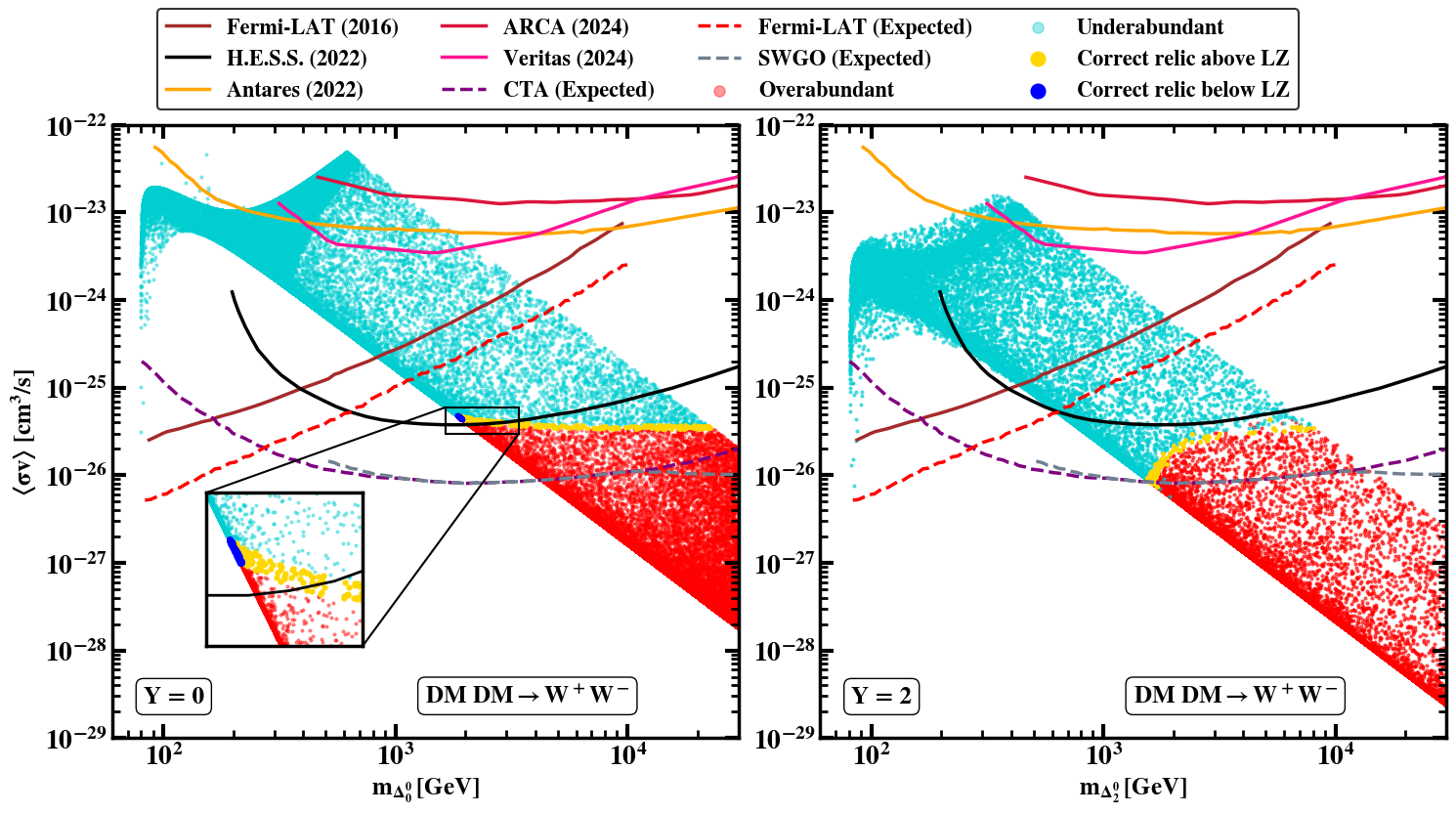}
    \caption{This figure shows the variation of thermally averaged annihilation cross-section
    $\langle \sigma v \rangle$ with mass of the scalar triplet DM candidate.
    The color scheme used is the same as in the above figures.
    The following experimental bounds have been added:
    CTA~\cite{CTAConsortium:2017dvg},
    Fermi-LAT~\cite{Fermi-LAT:2016uux},
    HESS~\cite{HESS:2022ygk},
    Antares and ARCA~\cite{KM3NeT:2024xca},
    Veritas~\cite{VERITAS:2024usn},
    SWGO~\cite{Viana:2019ucn}.}
    \label{fig:sigmav2}
\end{figure}
Fig.~\ref{fig:sigmav2} shows the thermally averaged annihilation
cross-section $\langle\sigma v\rangle$ into the $W^{+}W^{-}$ final state as a
function of the DM mass for the scalar triplet with hypercharge $Y = 0$ (left
panel) and $Y = 2$ (right panel), following the same color convention as before.
We focus on the $W^{+}W^{-}$ channel, as it is the dominant annihilation mode mediated by electroweak gauge interactions over a wide range of parameter space.
In both scenarios, the annihilation is dominated by electroweak gauge
interactions, yielding a cross-section that decreases approximately as
$m_{\Delta}^{-2}$ in the heavy-mass regime.  The sharp enhancement in
$\langle\sigma v\rangle$ near $m_{\Delta_{0(2)}^0} \simeq m_{W}$ corresponds to the
kinematic threshold for on-shell $W^{+}W^{-}$ production.  Overlaid on each
panel are the current experimental upper limits and projected sensitivities from
Fermi-LAT~\cite{Fermi-LAT:2016uux}, H.E.S.S.~\cite{HESS:2022ygk},
ANTARES and ARCA~\cite{KM3NeT:2024xca}, VERITAS~\cite{VERITAS:2024usn},
CTA~\cite{CTAConsortium:2017dvg}, and SWGO~\cite{Viana:2019ucn}. \\
For the $Y = 0$ case (left panel), the annihilation is driven purely by $SU(2)_{L}$ gauge interactions, yielding a characteristic cross-section of $\langle\sigma v\rangle \sim 10^{-24}$--$10^{-25}~\mathrm{cm^{3}\,s^{-1}}$ in the TeV mass range.  Upon imposing simultaneously the observed relic abundance, the direct detection bound from LZ, and the H.E.S.S.\ upper limit~\cite{HESS:2022ygk}, no viable points survive in the $Y = 0$ parameter space, indicating that the combination of current direct and indirect detection constraints are already sufficient to completely exclude this scenario in the mass region where the relic density is correctly reproduced. \\
For the $Y = 2$ case (right panel), which is already entirely ruled out by 
direct detection due to $Z$-boson mediated spin-independent scattering, the 
correct-relic points above the LZ bound appear at 
$m_{\Delta_2^0} \gtrsim \mathcal{O}(1)$~TeV with thermally averaged cross 
sections $\langle \sigma v \rangle \sim 10^{-26}~\text{cm}^3/\text{s}$. These 
points lie below the current H.E.S.S. sensitivity but fall within the 
projected reach of CTA and SWGO, indicating that while this scenario is 
disfavoured by direct searches, the indirect detection channel remains a 
complementary probe of the high-mass region.
 Hence looking ahead to future sensitivities, the projected CTA limits are expected to
probe and constrain even more, the already ruled out $Y = 0$ and $Y = 2$ scalar triplet scenarios. 
Thus, to conclude, both $Y=0$ and $Y=2$ scalar triplet DMs are completely ruled out by the current experimental data.

\subsection{Fermion Triplet DM} \label{subsec:fermiontriplet}

We now discuss the DM phenomenology of the fermion triplet, which
presents a considerably simpler parameter space than its scalar counterpart.
Unlike the scalar triplet, the fermion triplet does not admit a renormalisable Higgs portal interaction. Consequently, there are no additional scalar couplings to vary, and the relic density and detection prospects are governed solely by the triplet fermion mass and its gauge interaction with the SM gauge bosons. The annihilation and co-annihilation processes of the fermion triplet are governed predominantly by electroweak gauge interactions mediated by the $W$ and $Z$ boson, rendering the scenario highly predictive, with the annihilation cross-section largely determined by the triplet mass. The triplet fermion mass is scanned over the
same range as adopted for the scalar triplet, $m_{\Sigma^0} \in [10^{0},\,
3\times 10^{4}]$~GeV, for direct comparability between the two cases.  As a
result, for each hypercharge assignment the relic density, direct detection
signal, and indirect detection cross-section are all uniquely determined once
$m_{\Sigma^0}$ is fixed, rendering the fermion triplet one of the most
constrained and falsifiable realisations of electroweak multiplet DM. We now discuss the relic density behaviour of the fermion triplet for both $Y=0$ and $Y=2$ scenarios.
\subsubsection{Relic density} \label{subsub:relicFT}
\begin{figure}[H]
    \centering
    \includegraphics[width=1\linewidth]{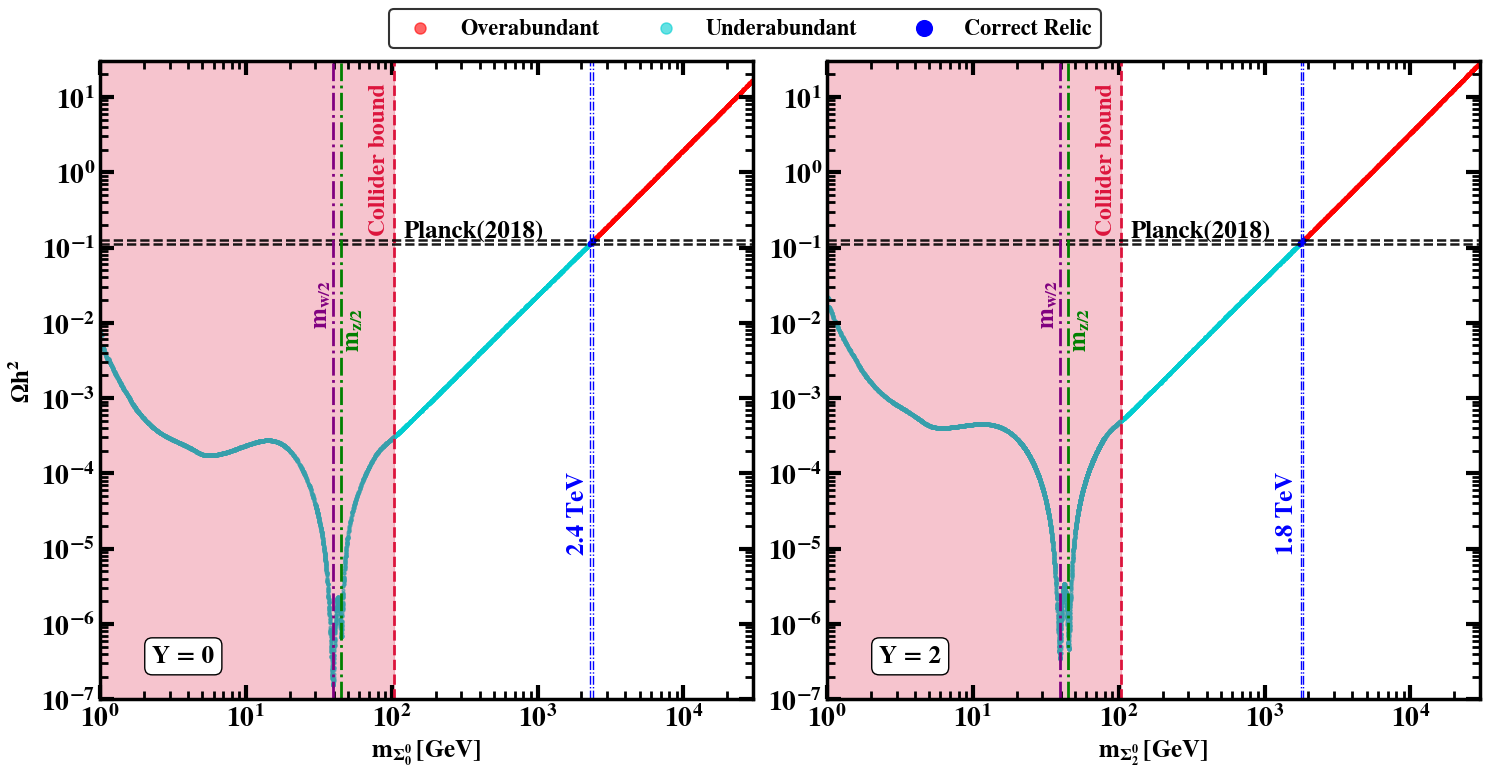}
    \caption{Relic density as a function of triplet fermion DM mass. The color scheme is same as that used in the scalar DM case.}
    \label{fig:relicFT}
\end{figure}

 We present the analysis of the fermion triplet DM scenario, beginning with the dependence of the relic density on the mass of its neutral component, as illustrated in the left (right) panel of Fig.~\ref{fig:relicFT} for the hypercharge $Y=0$ ($Y=2$) case.
The mass splitting between the neutral and charged components is 160 MeV for $Y=0$. In the $Y=2$ case, the splitting is 160 MeV between the singly charged and neutral components and 675 MeV between the doubly charged and neutral components \cite{McKay:2017xlc}. The colour scheme is the same as before. The low-mass region below $103.5$ GeV, indicated by the crimson-shaded area, is ruled out by LEP  constraints~\cite{ALEPH:2003acj,DELPHI:2003uqw,L3:1999onh, 
vonderPahlen:2016cbw}. The imposed $\mathbb{Z}_2$ symmetry forbids the $\bar{L}_{L}\tilde{H}\Sigma$ interaction in the pure fermion triplet DM scenario, thereby eliminating any Higgs-mediated resonance. In analogy with the scalar case, the low-mass region exhibits a band that shifts upward with increasing mass splitting. \\
We observe characteristic dips at $\sim 40.2~\mathrm{GeV}$ and $\sim 45.6~\mathrm{GeV}$, corresponding to approximately half the $W$ and $Z$ boson masses, respectively. While both features arise from $s$-channel resonant enhancement, the dip at $\sim 40.2~\mathrm{GeV}$ is associated with processes mediated by the $W$-boson, whereas the dip at $\sim 45.6~\mathrm{GeV}$ originates from the $\Sigma^{+}\Sigma^{-}$ coannihilation channel(as shown in Fig.\ref{fig:RelicTF0}) through a $s$-channel $Z$-boson. The blue points indicate a narrow mass region of 2.29–2.41 TeV for $Y=0$, while for $Y=2$ they lie in the range 1.78–1.83 TeV. The relic density is achieved at a lower mass in the $Y=2$ case compared to $Y=0$, due to enhanced annihilation driven by additional hypercharge interactions and stronger coannihilation effects involving multiple charged states. We now discuss the direct detection prospects of fermion triplet DM.

\subsubsection{Direct Detection} \label{subsubsec:DDFT}

\begin{figure}[H]
    \centering
    \includegraphics[width=0.5\linewidth]{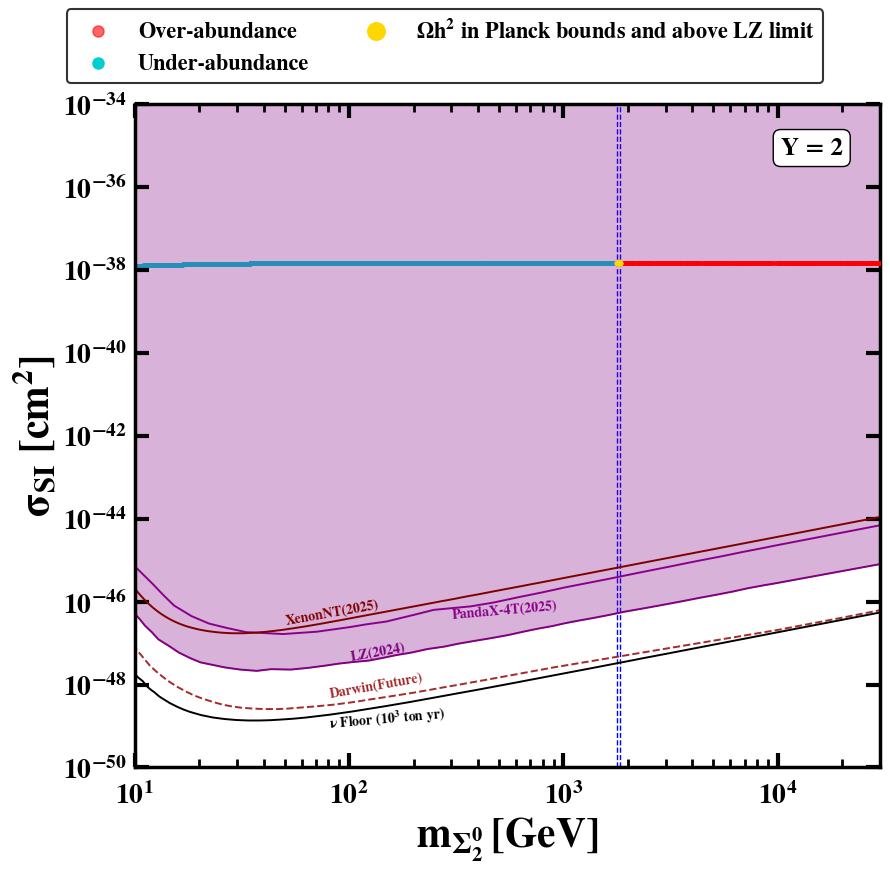}
    \caption{\small Spin-independent direct detection cross-section for Y = 2 fermion triplet DM case. The analogous DM-nucleon cross-section is absent for $Y=0$ case.}
    \label{fig:FDDY2}
\end{figure}

For the fermion triplet with $Y=0$, the spin-independent and spin-dependent direct detection cross-sections vanish at tree level. This is due to the absence of Higgs and $Z$ couplings for the neutral component, while $W$-mediated interactions are inelastic and kinematically suppressed. Consequently, only loop-induced processes generate extremely small scattering rates. In contrast, for the fermion triplet with $Y=2$, spin-independent scattering arises at tree level via vector interactions mediated by the $Z$-boson, leading to a sizable direct detection cross-section. As a result, stringent constraints from current experiments place strong bounds on the viable parameter space. In addition, spin-dependent scattering is also present due to the axial-vector coupling of the DM to the $Z$-boson. Although the spin-dependent cross-section can be significant, the corresponding experimental limits are comparatively weaker, and thus the direct detection constraints are primarily driven by the spin-independent interaction. Fig.~\ref{fig:FDDY2} shows the spin-independent direct detection cross-section for $Y=2$ fermion triplet DM case. We find a spin-independent cross-section of $\sim 10^{-38}$ cm$^2$ in our scan. This value is several orders of magnitude above the current experimental upper bounds, 
implying that a Dirac fermion triplet with $Y=2$ is excluded\footnote{Potential way out will be to suppress the 
$Z$-mediated interaction  e.g. by splitting the neutral state into 
Majorana components or introducing inelastic scattering.}
In conclusion, for $Y=0$ the absence of a tree-level $Z$ coupling leads to a highly suppressed spin-independent cross-section, whereas for $Y=2$ the presence of a $Z$-mediated interaction results in a large cross-section that is strongly constrained and typically excluded by current direct detection experiments. We now proceed to examine the indirect detection prospects of fermion triplet DM.

\subsubsection{Indirect Detection} \label{subsubsec:IDFT}
\begin{figure}[H]
    \centering
    \includegraphics[width=1\linewidth]{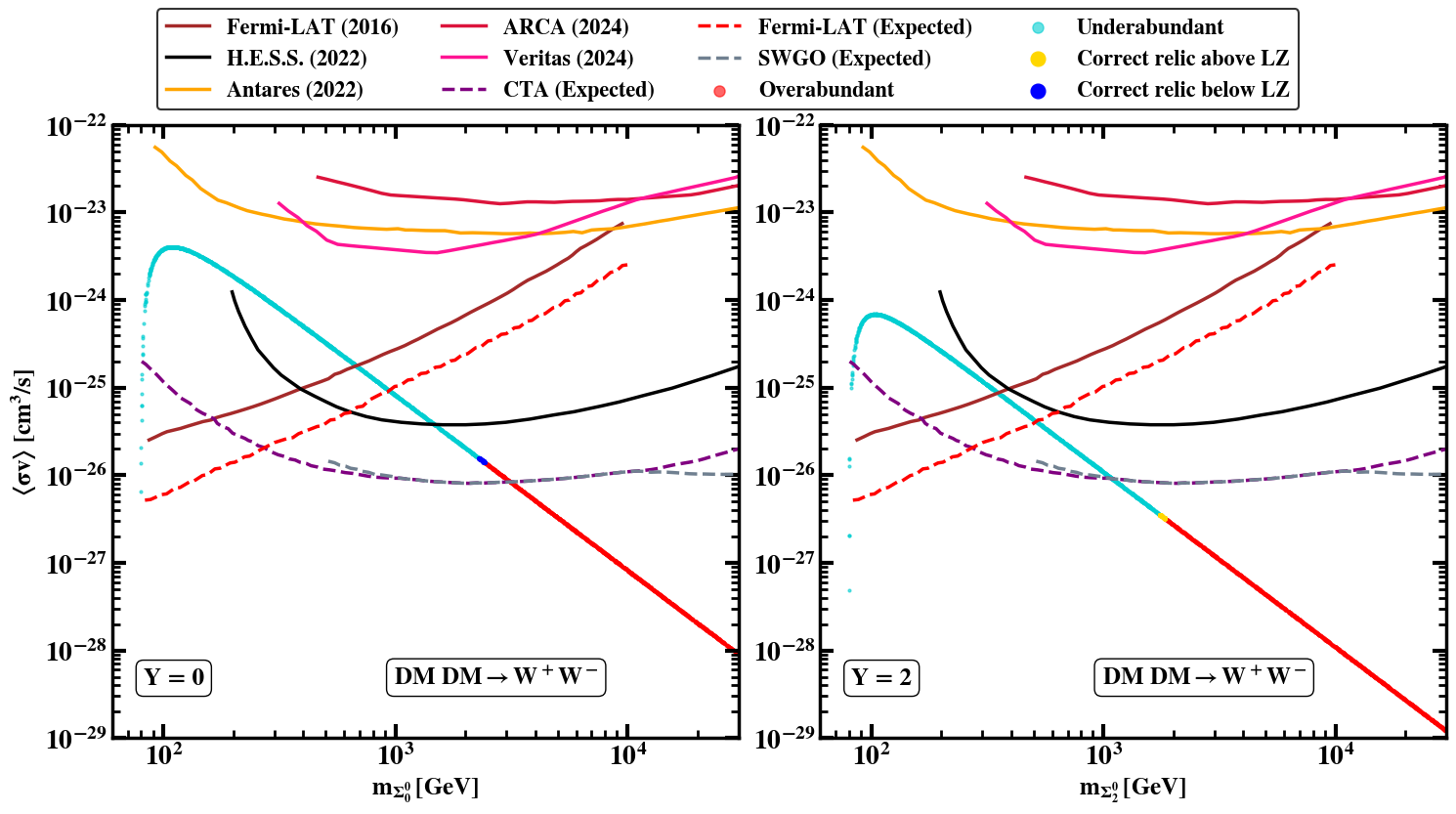}
    \caption{\small
    This figure shows the variation of the thermally averaged annihilation cross-section $\langle \sigma v \rangle$
    with the mass of fermion triplet DM candidate $Y=0(Y=2)$ in the left (right) panel.
    The color scheme used is the same as in the above figures. The lines correspond to the following current and upcoming experimental bounds:
    CTA~\cite{CTAConsortium:2017dvg},
    Fermi-LAT~\cite{Fermi-LAT:2016uux},
    HESS~\cite{HESS:2022ygk},
    Antares and ARCA~\cite{KM3NeT:2024xca},
    Veritas~\cite{VERITAS:2024usn},
    SWGO~\cite{Viana:2019ucn}.}
    \label{fig:FIDY2}
\end{figure}

 To quantify the indirect detection signal, we compute the thermally averaged DM annihilation cross-section with the $W^{+}W^{-}$ channel providing the dominant contribution. Fig.~\ref{fig:FIDY2} displays the thermally averaged annihilation
cross-section $\langle\sigma v\rangle$ into the $W^{+}W^{-}$ final state as a
function of the DM mass for the fermion triplet with hypercharge
$Y = 0$ (left panel) and $Y = 2$ (right panel). The dominant contribution in
both cases arises from electroweak gauge interactions, leading to a characteristic
behaviour where the cross-section decreases with increasing mass in the
heavy-mass regime, with a mild enhancement near the kinematic threshold
$m_{\Sigma_{0(2)}^0} \sim m_{W}$. The current experimental limits and projected
sensitivities are also shown for comparison. 
A striking feature of the fermion triplet scenario is that the allowed parameter
space is confined to a narrow trajectory rather than forming an extended band.
This reflects the highly constrained nature of the model, where the annihilation
process is almost entirely governed by gauge interactions, leaving little freedom
to vary the cross-section independently of the DM mass. Consequently,
imposing the relic density condition effectively fixes the mass of the DM candidate, resulting in a sharply defined relation between
$\langle\sigma v\rangle$ and $m_{\Sigma_{0(2)}^0}$. \\
The correct relic density region appears as a narrow band of points, which remains largely unconstrained by current indirect detection experiments for $Y=0$ triplet fermion DM, whereas in the $Y=2$ scenario, the relic density-favoured region remains beyond the reach of both current and projected experimental sensitivities. This sharply defined structure highlights the predictive nature of the fermion triplet model, in contrast to the scalar case where additional interactions introduce a broader spread in the parameter space. \\
Overall, the fermion triplet DM model exhibits distinct phenomenological features for the two hypercharge assignments $Y = 0$ and $Y = 2$. In the $Y = 0$ scenario, both spin-independent and spin-dependent direct detection cross-sections vanish at leading order, resulting in no observable signal, and current indirect detection constraints do not exclude this case, although future experiments such as CTA and SWGO will potentially rule out the entire viable parameter space. In contrast, the $Y = 2$ scenario is already excluded by current direct detection experiments, although both current and projected indirect detection limits remain insufficient to constrain it. In summary, for $Y=2$ case, the entire parameter space is already ruled out by the direct detection experiments. In contrast, for $Y=0$ case, there is no constraint from direct detection experiments  while the current indirect detection bounds are weak. However, upcoming indirect detection experiments will likely rule out the entire viable parameter space.

\section{Conclusions}
\label{sec:conclusions}

 We have investigated the DM phenomenology of electroweak triplet 
extensions of the SM, considering both scalar and fermionic candidates with 
hypercharges $Y=0$ and $Y=2$. We have computed the relic abundance, direct 
detection cross-sections, and indirect detection signals across the relevant 
parameter space, incorporating the effects of electroweak gauge interactions 
and radiative mass splittings. Our results show that the scalar triplet with 
$Y=0$ is effectively ruled out by the combination of relic density, direct detection and indirect 
detection limits. The scalar triplet with $Y=2$ is entirely excluded by the combination of relic density and 
direct detection experiments due to its unsuppressed tree-level $Z$-boson 
exchange contribution to the spin-independent DM-nucleon scattering cross 
section, which yields $\sigma_{\rm SI} \sim 10^{-38}~\text{cm}^2$, exceeding 
the current LZ exclusion limit by several orders of magnitude. Among the fermionic cases, the triplet with $Y=2$ is also ruled out by the combination of relic density and direct detection bounds.
The remaining case, namely the fermionic triplet with $Y=0$ retains a viable parameter space consistent with current observations. \\
We find that the phenomenology is highly sensitive to the hypercharge 
assignment. In particular, the $Y=2$ scalar and fermion cases exhibit an enhanced 
coupling to the $Z$-boson arising from its non-zero hypercharge, which 
mediates unsuppressed spin-independent DM-nucleon scattering at tree level and renders 
the entire parameter space incompatible with current direct detection bounds. 
The $Y=0$ scalar scenario, while free from this constraint, is instead 
disfavoured by indirect detection limits, in particular from the H.E.S.S data. The current indirect detection limits do not rule out the $Y=0$ fermionic case. However, the surviving parameter space in the fermionic triplet models lies within the 
reach of future experiments, with indirect detection searches by CTA and SWGO 
expected to play a particularly important role in probing the high-mass region 
of these scenarios and will potentially rule out the entire parameter space.

\section*{Acknowledgments}

We are grateful to Anirban Majumdar for providing his (then unpublished) results on the limits associated with the neutrino floor and fog~\cite{DeRomeri:2025nkx}. The authors would like to acknowledge the use of \textit{FeynRules 2.3} \cite{Alloul:2013bka} for model implementation and \textit{MicrOMEGAs} 6.3.0 \cite{Belanger:2006is,Belanger:2014vza,Alguero:2023zol} for computing the thermal relic abundance of DM, as well as the DM-nucleon scattering and annihilation cross-sections.

\appendix

\section{Annihilation, Production and Detection of DM}
\label{sec:appendix}
\begin{figure}[H]
        \centering
    \includegraphics[height=13cm]{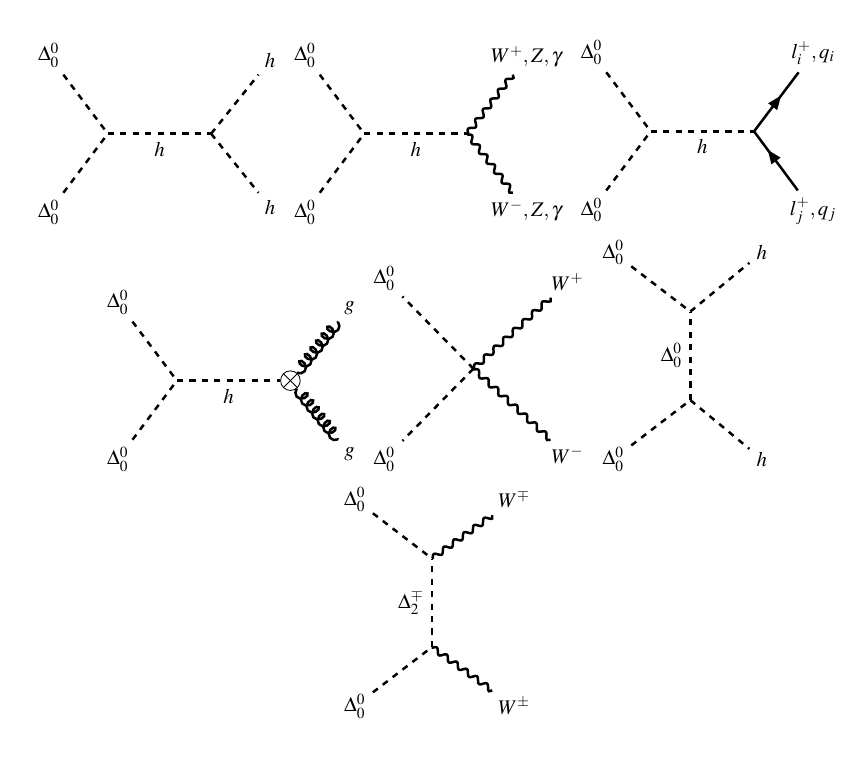}
        \caption{Relevant diagrams (annihilation channels) for computing the relic density for $Y=0$ scalar triplet.}
            \label{fig:annihiTS0}
\end{figure}

\begin{figure}[H]
        \centering
    \includegraphics[height=13cm]{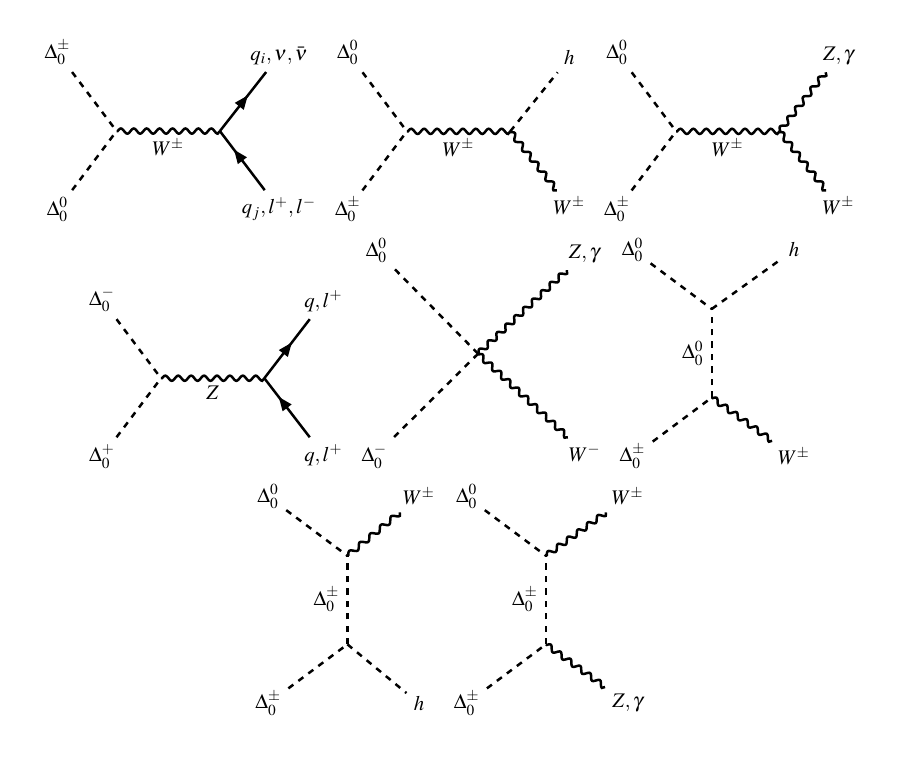}
        \caption{Relevant diagrams (co-annihilation channels) for computing the relic density for $Y=0$ scalar triplet.}
        \label{fig:CoannihiTS2_y0}
\end{figure}

\begin{figure}[H]
        \centering
    \includegraphics[height=20cm]{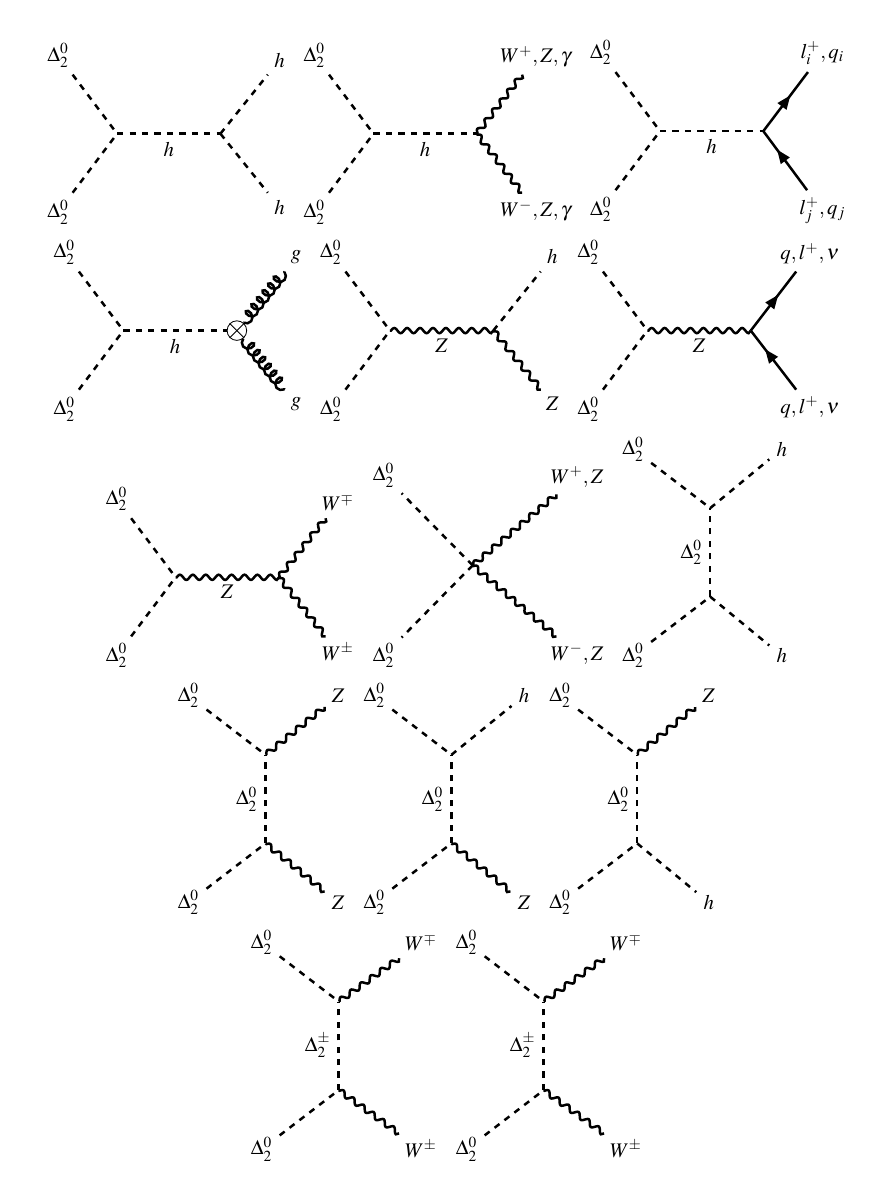}
        \caption{Relevant diagrams (annihilation channels) for computing the relic density for $Y=2$ scalar triplet.}
            \label{fig:annihiTS2}
\end{figure}

\begin{figure}[H]
        \centering
    \includegraphics[height=18cm]{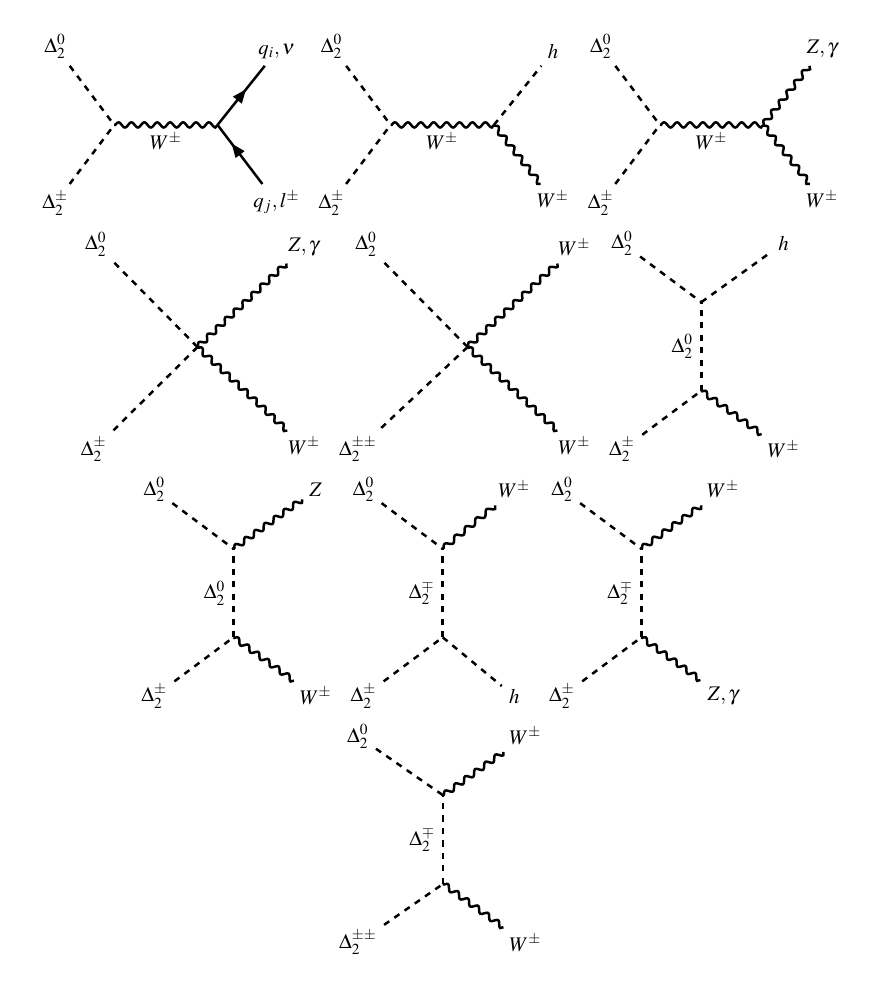}
        \caption{Relevant diagrams (co-annihilation channels) for computing the relic density for $Y=2$ scalar triplet.}
        \label{fig:CoannihiTS2}
\end{figure}

\begin{figure}[H]
        \centering
    \includegraphics[height=10cm]{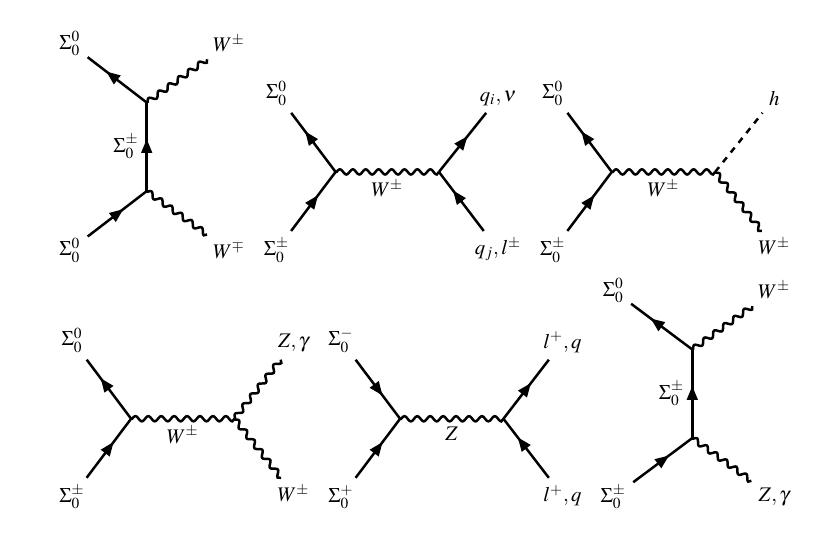}
        \caption{Relevant diagrams (annihilation and co-annihilation channels) for computing the relic density for $Y=0$ fermion triplet.}
          \label{fig:RelicTF0}
\end{figure}

\begin{figure}[H]
        \centering
    \includegraphics[height=18cm]{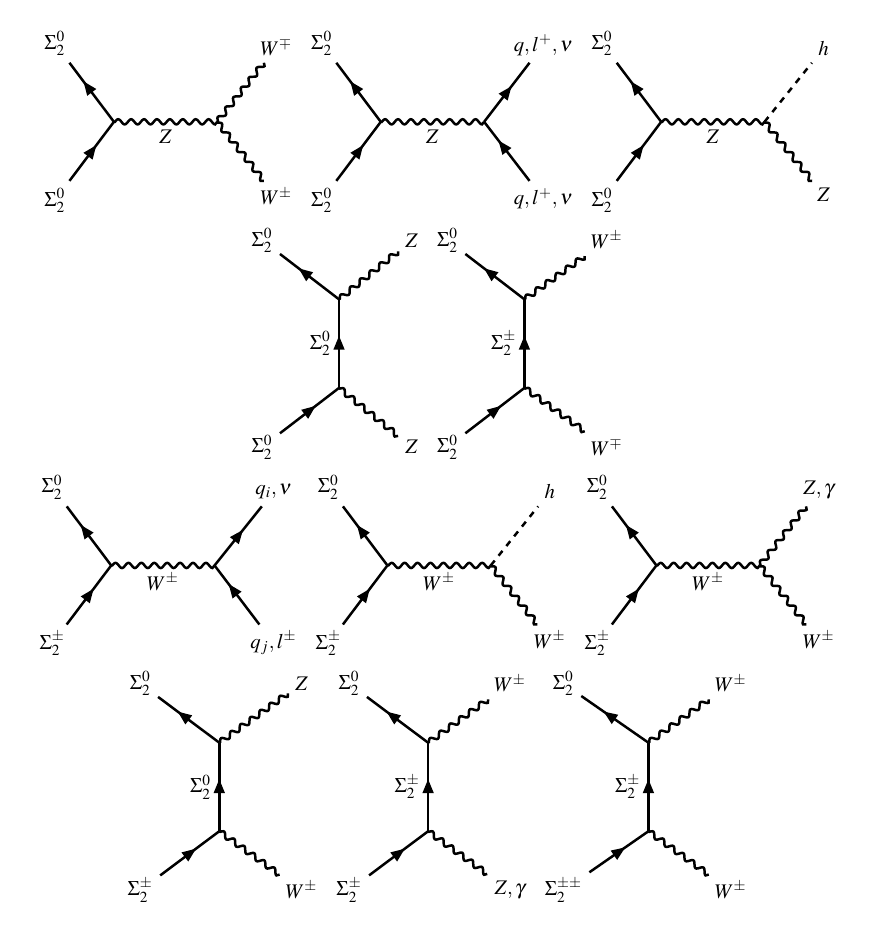}
        \caption{Relevant diagrams (annihilation and co-annihilation channels) for computing the relic density for $Y=2$ fermion triplet.}
            \label{fig:RelicTF2}
\end{figure}

\begin{figure}[H]
        \centering
    \includegraphics[height=5.0cm]{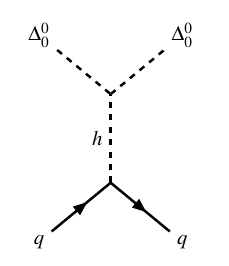}
     \includegraphics[height=5.0cm]{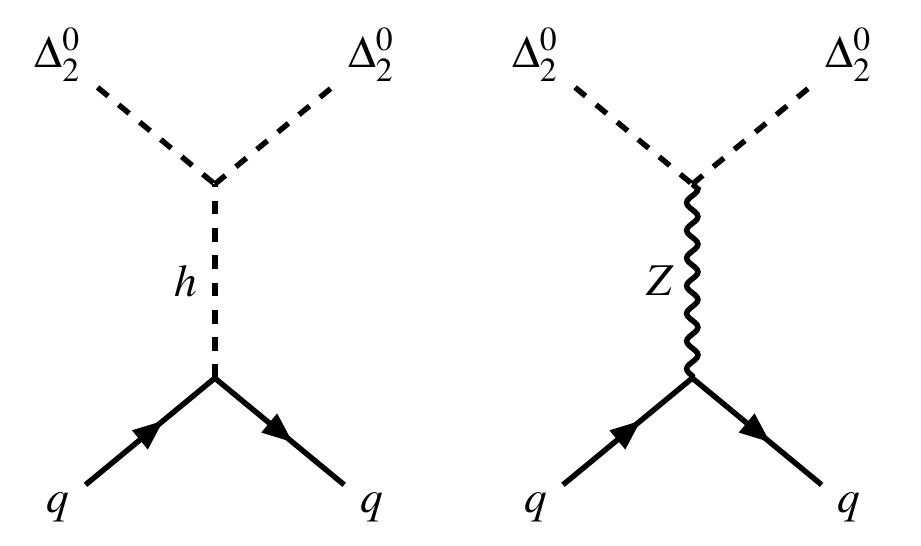}
        \caption{Relevant diagrams for the direct detection of the scalar triplet of hypercharge Y=0 and Y=2.}
           \label{fig:ddTS0} 
\end{figure}

\begin{figure}[H]
\centering
        \includegraphics[height=11cm]{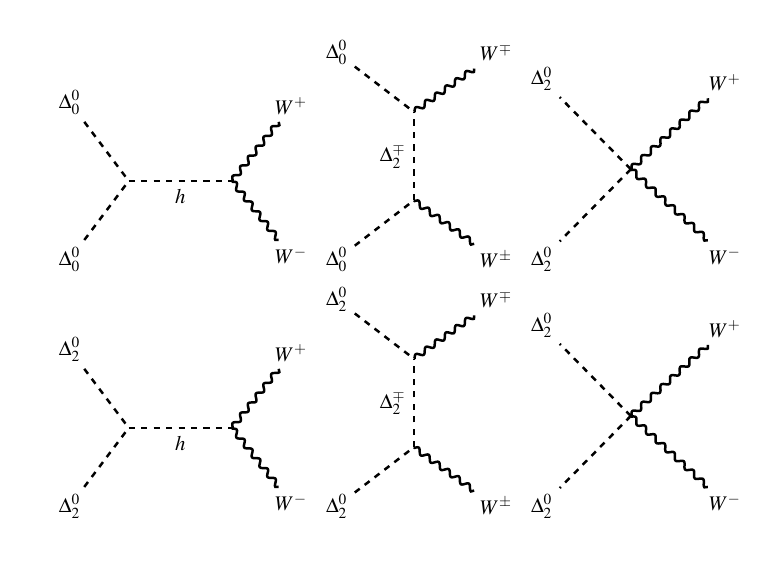}
        \caption{Dominant Feynman diagrams contributing to the indirect detection signals of the scalar triplet hypercharge Y=0 (upper row) and Y=2 (lower row).}
                \label{fig:ddTS01} 
\end{figure}

\begin{figure}
    \centering
    \includegraphics[height=4.8cm]{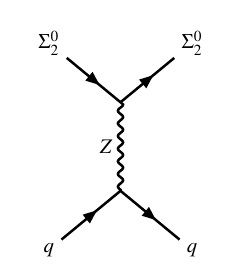}
    \caption{Relevant diagrams for the direct detection of the fermion triplet of hypercharge Y = 2.}
    \label{fig:FDD}
\end{figure}

\begin{figure}
    \centering
    \includegraphics[width=0.8\linewidth]{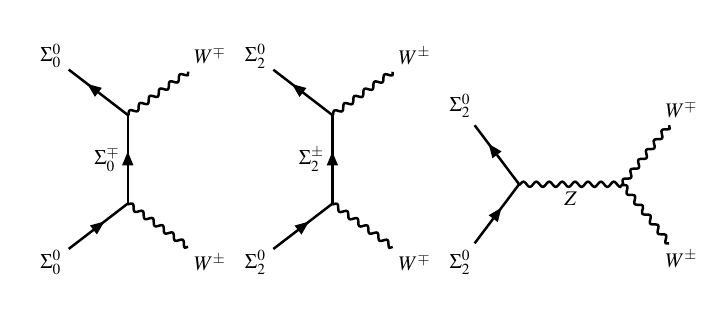}
    \caption{Dominant Feynman diagrams contributing to the indirect detection signals of the fermion triplet hypercharge Y=0 (left) and Y=2 (right).}
    \label{fig:ddTF01}
\end{figure}

 In Figs.~\ref{fig:annihiTS0} to \ref{fig:RelicTF2}, we list the possible diagrams for production/annihilation of DM, relevant in the early universe, for the cases in which the DM is scalar triplet or fermion triplet, respectively. In  Figs. \ref{fig:ddTS0} and \ref{fig:FDD}, we show the direct detection prospects of the scalar and fermion triplet DM by the exchange of a Higgs or $Z$ bosons. In Figs.~\ref{fig:ddTS01} and \ref{fig:ddTF01}, we present the indirect detection prospects for scalar and fermion triplet DM candidates, focusing on their annihilation into $W^{+}W^{-}$ final states.

\section{Collider Prospects of triplet DM}
\label{sec:Collider}
At colliders, the new scalar or fermion triplet fields can manifest through several distinctive features. Their production primarily occurs in pairs via electroweak Drell--Yan processes involving both charged and neutral currents. Once produced, these scalars decay through interactions with the Higgs and gauge sectors, giving rise to lighter scalar states together with on- or off-shell Higgs and weak gauge bosons. The presence of a conserved $\mathbb{Z}_2$ symmetry ensures that every production event ultimately contains the stable neutral particle i.e. the DM in the final state. As a result, collider events associated with this model typically exhibit large missing transverse momentum originating from the invisible DM escaping detection, accompanied by visible decay products such as leptons, jets, or photons from Higgs and gauge boson decays. \\
Such characteristic signatures are probed in multiple ongoing ATLAS and CMS searches targeting final states with missing energy and electroweak bosons. The scalar and fermion triplets predicted in our model can also be probed at future high-energy hadron colliders such as the Future Circular Collider (FCC-hh) \cite{FCC:2019vvp, Mangano:2017tke} and the Super Proton-Proton Collider (SPPC) \cite{CEPC-SPPCStudyGroup:2015csa, Helsens:2019ehy}, operating at centre-of-mass energies of up to 100 TeV. In Table~\ref{tab:triplet_signatures_clean}, we summarise the production channels, decay modes, and characteristic collider signatures of scalar and fermion triplets for both hypercharge assignments. In the following subsections, we briefly discuss the production prospects of the scalar triplet $(Y=2)$ and the fermionic triplet $(Y=0)$. The other two cases, scalar triplet with $Y=0$ and fermion triplet with $Y=2$ have similar signatures, which we do not discuss separately to avoid unnecessary repetition.
 
\begin{table}[H]
\centering
\small
\begin{tabular}{|l|p{8cm}|p{3.5cm}|}
\hline
\textbf{Production} & \textbf{Decay mode} & \textbf{Signature} \\
\hline\hline
\multicolumn{3}{|c|}{\textbf{Scalar triplet } $\boldsymbol{\Delta_{0}=\{\Delta_{0}^{\pm},\Delta_{0}^{0}\}}$} \\
\hline
$pp \to \Delta_{0}^{+}\Delta_{0}^{-}$ & $\Delta_{0}^{\pm} \to \Delta_{0}^{0} W^{\pm(*)} \to \Delta_{0}^{0} \ell^{\pm} \nu$ & $2\ell + \slashed{E}_T$ \\
$pp \to \Delta_{0}^{\pm}\Delta_{0}^{0}$ & $\Delta_{0}^{\pm} \to \Delta_{0}^{0} W^{\pm(*)} \to \Delta_{0}^{0} \ell^{\pm} \nu$ & $1\ell + \slashed{E}_T$ \\
$e^{+}e^{-} \to \Delta_{0}^{+}\Delta_{0}^{-}$ & $\Delta_{0}^{\pm} \to \Delta_{0}^{0} W^{\pm(*)} \to \Delta_{0}^{0} \ell^{\pm} \nu$ & $2\ell + \slashed{E}_T$ \\
\hline
\multicolumn{3}{|c|}{\textbf{Scalar triplet } $\boldsymbol{\Delta_{2}=\{\Delta_{2}^{\pm\pm},\Delta_{2}^{\pm},\Delta_{2}^{0}\}}$} \\
\hline
$pp \to \Delta_{2}^{++}\Delta_{2}^{--}$ & $\Delta_{2}^{\pm\pm} \to \ell^{\pm}\ell^{\pm}$ & $4\ell$ + $\slashed{E}_T$ \\
$pp \to \Delta_{2}^{\pm\pm}\Delta_{2}^{\mp}$ & $\Delta_{2}^{\pm\pm} \to \Delta_{2}^{\pm} W^{\pm(*)} \to \Delta_{2}^{0} W^{\pm(*)} W^{\pm(*)}$ & $3\ell + \slashed{E}_T$ \\
$e^{+}e^{-} \to \Delta_{2}^{++}\Delta_{2}^{--}$ & $\Delta_{2}^{\pm\pm} \to \ell^{\pm}\ell^{\pm}$ & $4\ell$ + $\slashed{E}_T$ \\
\hline
\multicolumn{3}{|c|}{\textbf{Fermion triplet } $\boldsymbol{\Sigma_{0}=\{\Sigma_{0}^{\pm},\Sigma_{0}^{0}\}}$} \\
\hline
$pp \to \Sigma_{0}^{+}\Sigma_{0}^{-}$ & $\Sigma_{0}^{\pm} \to \Sigma_{0}^{0} W^{\pm(*)} \to \Sigma_{0}^{0} \ell^{\pm} \nu$ & $2\ell + \slashed{E}_T$ \\
$pp \to \Sigma_{0}^{\pm}\Sigma_{0}^{0}$ & $\Sigma_{0}^{\pm} \to \Sigma_{0}^{0} W^{\pm(*)} \to \Sigma_{0}^{0} \ell^{\pm} \nu$ & $1\ell + \slashed{E}_T$ \\
$e^{+}e^{-} \to \Sigma_{0}^{+}\Sigma_{0}^{-}$ & $\Sigma_{0}^{\pm} \to \Sigma_{0}^{0} W^{\pm(*)} \to \Sigma_{0}^{0} \ell^{\pm} \nu$ & $2\ell + \slashed{E}_T$ \\
$pp, e^{+}e^{-} \to \Sigma_{0}^{\pm}\Sigma_{0}^{0}$ & $\Sigma_{0}^{0} \to W^{\pm}\ell^{\mp},~Z\nu,~h\nu$ & $3\ell + \slashed{E}_T$ \\
\hline
\multicolumn{3}{|c|}{\textbf{Fermion triplet } $\boldsymbol{\Sigma_{2}=\{\Sigma_{2}^{\pm\pm},\Sigma_{2}^{\pm},\Sigma_{2}^{0}\}}$} \\
\hline
$pp \to \Sigma_{2}^{++}\Sigma_{2}^{--}$ & $\Sigma_{2}^{\pm\pm} \to \ell^{\pm}\ell^{\pm}$ & $4\ell$ + $\slashed{E}_T$ \\
$pp \to \Sigma_{2}^{\pm\pm}\Sigma_{2}^{\mp}$ & $\Sigma_{2}^{\pm\pm} \to \Sigma_{2}^{\pm}W^{\pm(*)} \to \Sigma_{2}^{0} W^{\pm(*)}W^{\pm(*)}$ & $3\ell + \slashed{E}_T$ \\
$e^{+}e^{-} \to \Sigma_{2}^{++}\Sigma_{2}^{--}$ & $\Sigma_{2}^{\pm\pm} \to \ell^{\pm}\ell^{\pm}$ & $4\ell$ + $\slashed{E}_T$ \\
\hline
\end{tabular}
\caption{Summary of representative production and decay modes of scalar ($\Delta_Y$) and fermion ($\Sigma_Y$) triplets at hadron and lepton colliders, with their characteristic final-state signatures.}
\label{tab:triplet_signatures_clean}
\end{table}

\subsection*{Production of triplet scalar particles (Y=2)}

In this subsection, we discuss the collider production prospects of the scalar triplet with hypercharge $Y=2$. 
\begin{figure}[H] \centering
        \includegraphics[height=8cm]{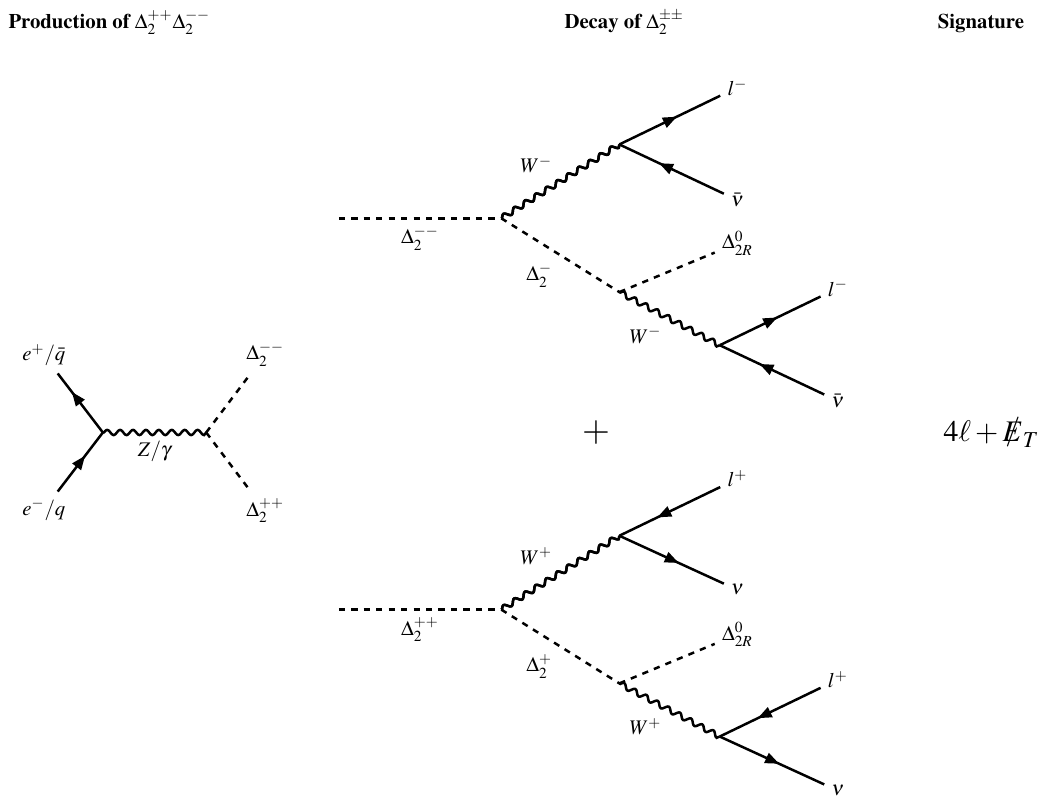}
        \includegraphics[height=8cm]{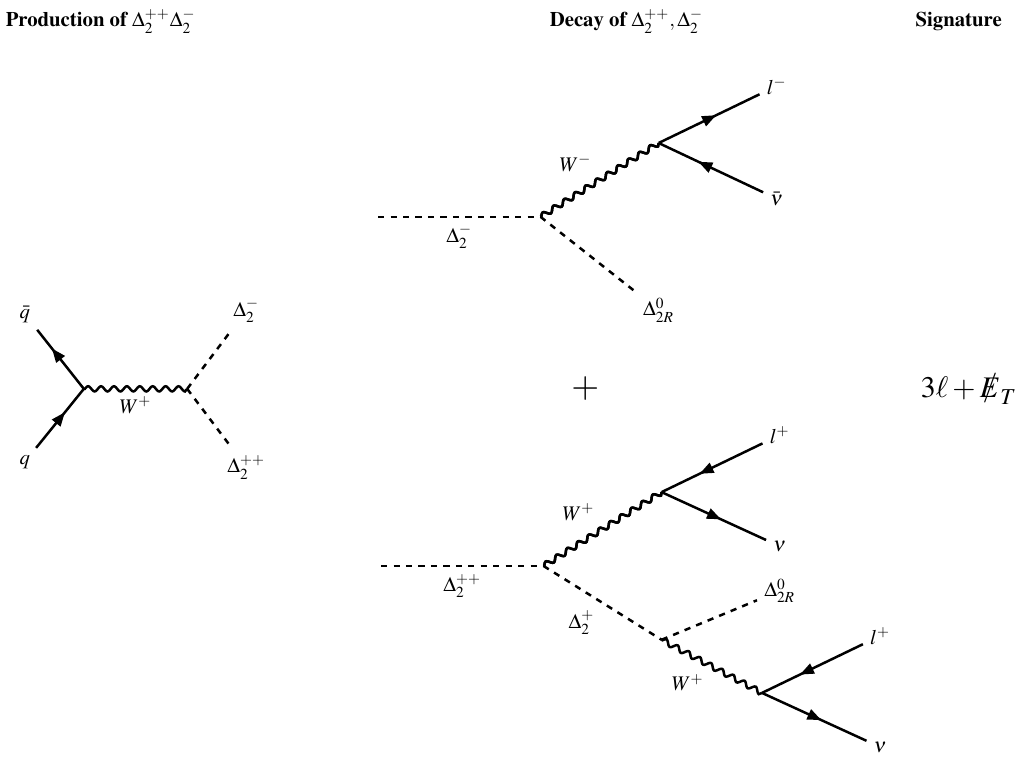}
        \caption{Illustrative representation of the doubly charged triplet scalar production in association with singly
charged triplet scalar and the expected decay channels for the case $m_{\Delta^{0}_{0}}<m_{\Delta^{-}_{0}}$.}
                \label{fig:Scalar2_Collider1}
\end{figure}

\begin{figure}[H]
    \centering
\includegraphics[height=8cm]{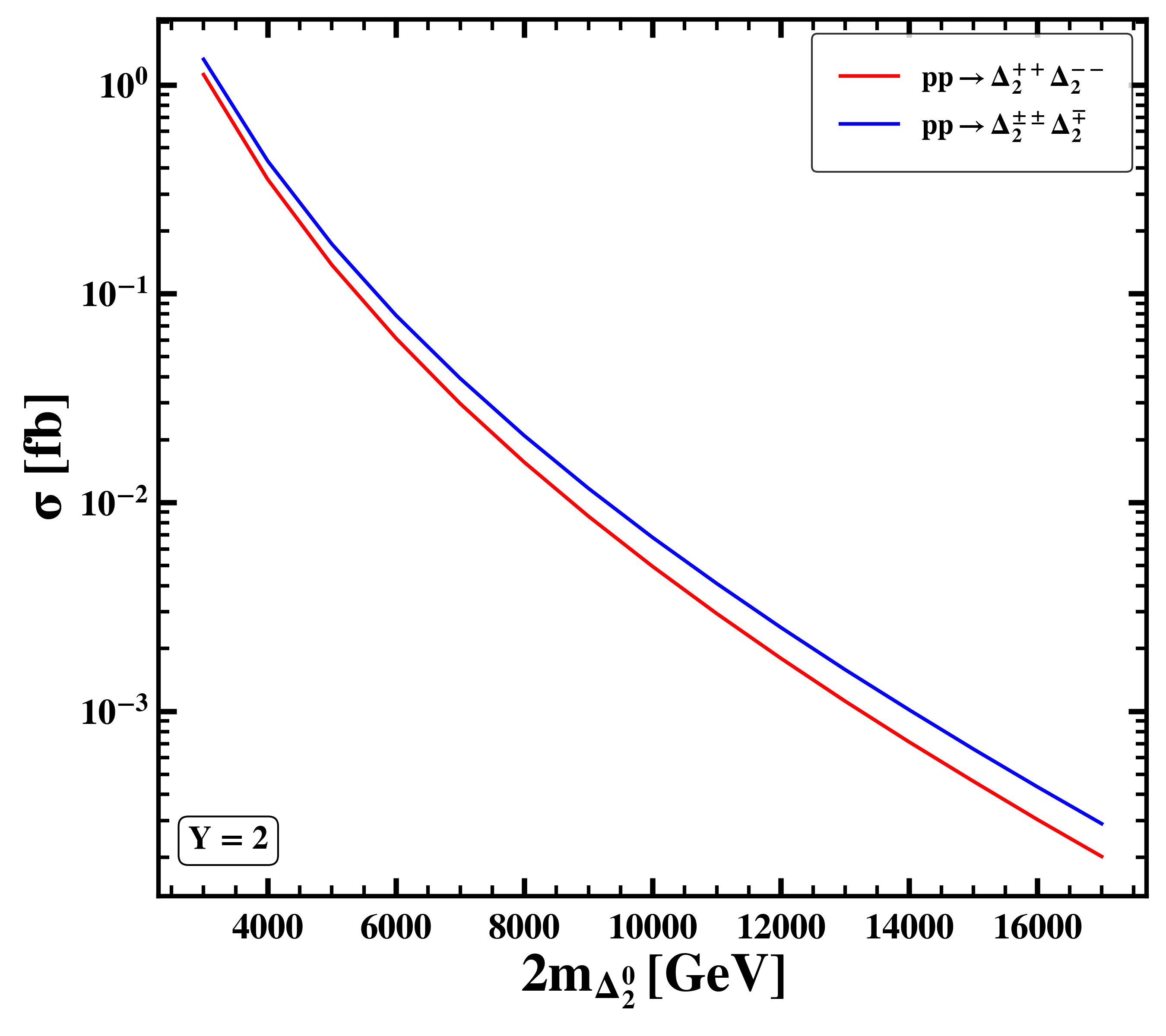}
    \caption{Production cross-section of triplet scalar $(Y=2)$ through Drell-Yan process at FCC-hh with $\sqrt{s}$ = 100 TeV. }
     \label{fig:Scalar2_Collider22}
\end{figure}

Being an electroweak multiplet, the triplet components are predominantly produced via gauge interactions, making the production cross-section largely independent of the scalar potential parameters. At hadron colliders such as the LHC, the dominant production channels arise from Drell–Yan processes mediated by $s$-channel $Z$/$\gamma$ and $W^{\pm}$ exchange, leading to 
$p p \rightarrow \Delta_2^{++}\Delta_2^{--}$, $p p \rightarrow \Delta_2^{\pm\pm}\Delta_2^{\mp}$, and $p p \rightarrow \Delta_2^{+}\Delta_2^{-}$. \\
In addition, at lepton colliders, the triplet scalars can be pair-produced via $s$-channel $Z$/$\gamma$ exchange, leading to $e^{+} e^{-} \rightarrow \Delta_2^{++}\Delta_2^{--}$ and $e^{+} e^{-} \rightarrow \Delta_2^{+}\Delta_2^{-}$. Subleading contributions may also arise from Higgs-mediated processes, such as gluon fusion via an off-shell Higgs boson or vector boson fusion channels involving Higgs exchange. However, these processes are typically suppressed by small scalar mixing and/or the off-shell Higgs propagator and hence contribute negligibly compared to the gauge-induced Drell–Yan production in most of the parameter space. In Fig.~\ref{fig:Scalar2_Collider1}, we present the Feynman diagrams for the production of the scalar triplet. In Fig.~\ref{fig:Scalar2_Collider22}, we present the production cross-section of triplet scalar $(Y = 2)$ through the Drell-Yan process at
FCC-hh with $\sqrt{s}$ = 100 TeV.

\subsection*{Production of triplet fermion particles (Y=0)}

\begin{figure}[H] \centering
        \includegraphics[height=8cm]{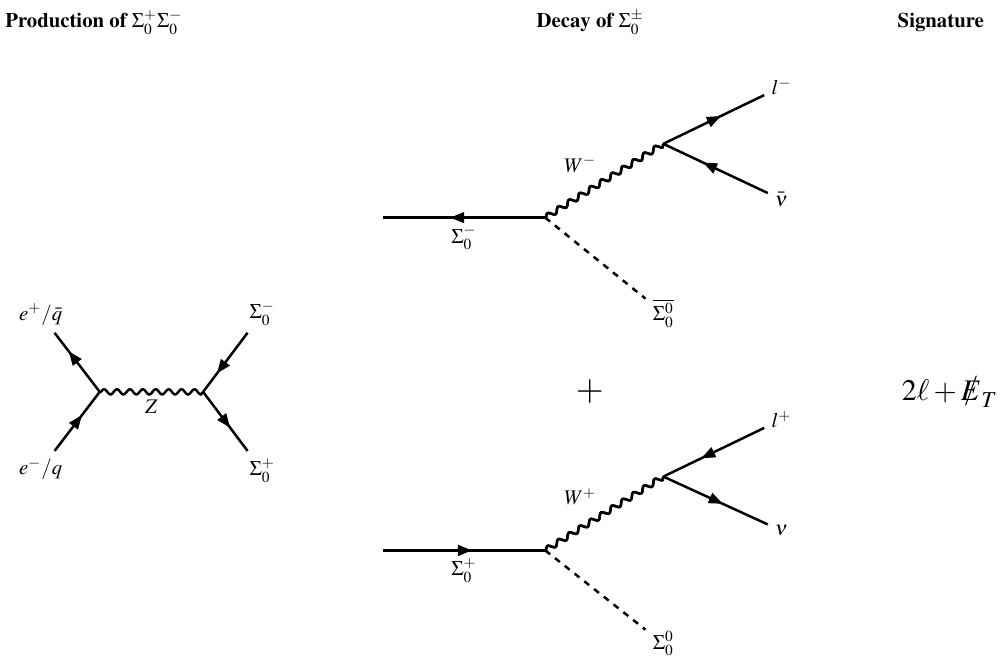}
        \caption{Illustrative representation of singly charged triplet fermion pair-production and decay channels.}
                \label{fig:Ferm0_Collider4}
\end{figure}

\begin{figure}[H]
    \centering
\includegraphics[height=8cm]{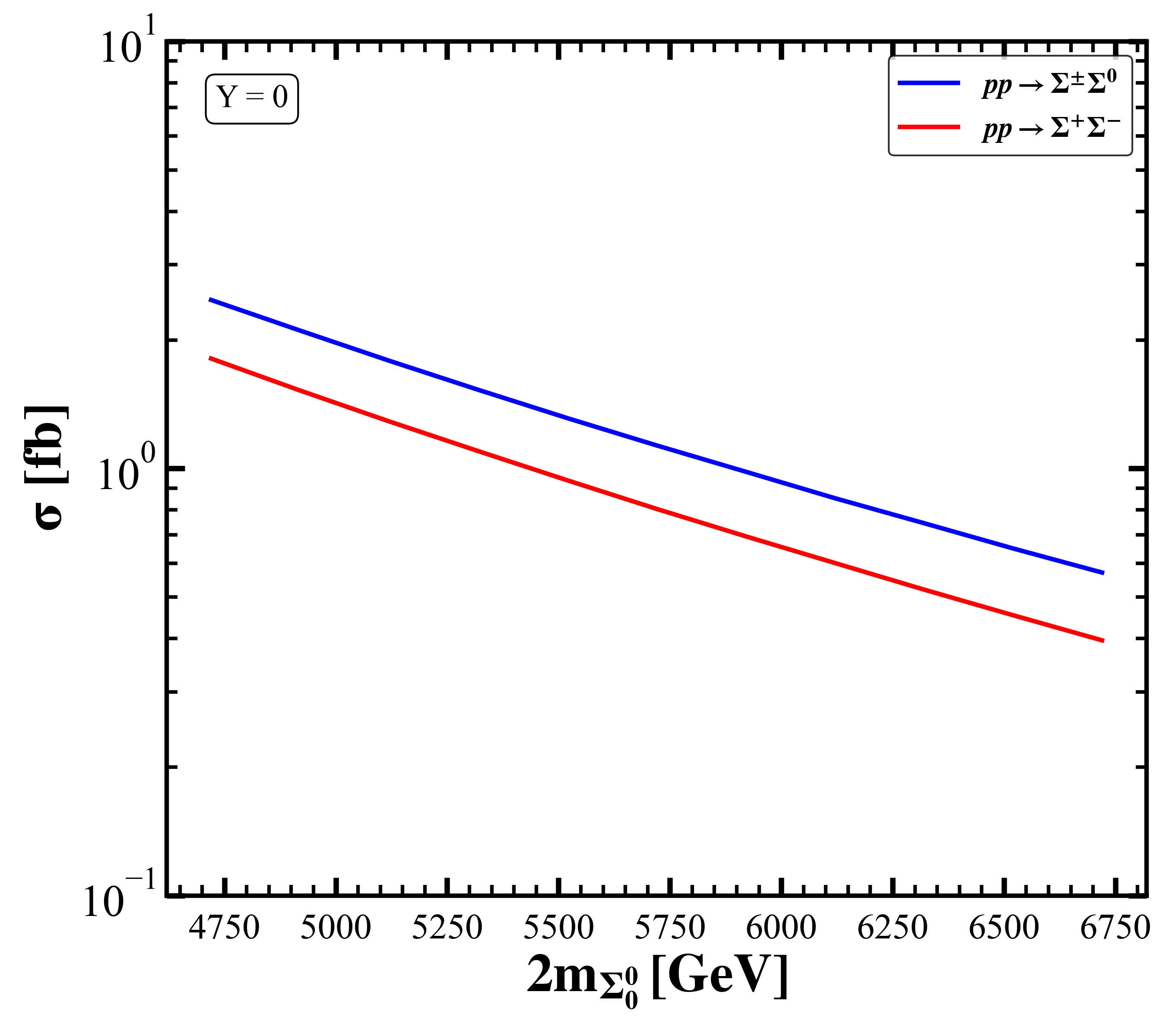}
    \caption{Production cross-section of triplet fermion $(Y=0)$ through Drell-Yan process at FCC-hh with $\sqrt{s}$ = 100 TeV.}
    \label{fig:Ferm0_cross-section}
\end{figure}

 Following the discussion of scalar triplet production, we now turn to the collider production prospects of the fermionic triplet with hypercharge $Y=0$. Similar to the scalar case, the production of the fermion triplet is governed predominantly by electroweak gauge interactions, rendering the leading-order cross-sections largely model-independent. At hadron colliders, the dominant production channels arise from Drell–Yan processes mediated by $s$-channel $Z$/$\gamma$ and $W^{\pm}$ exchange, leading to pair and associated production of the triplet components:
$p p \rightarrow \Sigma_0^{+} \Sigma_0^{-}$, $p p \rightarrow \Sigma_0^{\pm} \Sigma_0^{0}$. Owing to the Majorana nature of the neutral component, $\Sigma_0^{0}$ does not couple to the photon, and its production proceeds primarily via charged current interactions. Among these, the associated production $p p \rightarrow \Sigma_0^{\pm} \Sigma_0^{0}$ typically yields the largest cross-section due to the charged current interaction mediated by the $W^{\pm}$ boson. In Figs.~\ref{fig:Ferm0_Collider4} we present the Feynman diagrams for the production of the scalar triplet components. In Fig.~\ref{fig:Ferm0_cross-section}, we present the production cross-section of the components of the triplet fermion $(Y = 0)$ through the Drell-Yan and associated production process at
FCC-hh with $\sqrt{s}$ = 100 TeV.

\bibliography{bibliography.bib}
\bibliographystyle{utphys}
\end{document}